\documentclass[transmag]{IEEEtran}
\usepackage{latexsym}
\usepackage{hyperref}

\newlength \figwidth
\usepackage{ucs}
\usepackage[utf8x]{inputenc}
\usepackage[cmex10]{amsmath}
\usepackage{cite, amsfonts, amssymb, amsthm, bm, bbm, graphicx, relsize, multirow, booktabs, tikz,soul}
\usepackage[american]{babel}
\usepackage[T1]{fontenc}
\usepackage{dblfloatfix}
\usepackage{algorithmic, algorithm}
\usepackage[multiple]{footmisc}
\usepackage{tabulary}
\usepackage{enumitem}
\usepackage[font={small}]{caption}
\usepackage{subcaption}
\usepackage{float}
\theoremstyle{definition}

\theoremstyle{remark}

\newcommand{\ds}{\displaystyle}

\newcommand{\bbeta}{\boldsymbol{\eta}}

\def\BibTeX{{\rm B\kern-.05em{\sc i\kern-.025em b}\kern-.08em T\kern-.1667em\lower.7ex\hbox{E}\kern-.125emX}}
\begin{document}

\bstctlcite{IEEE_nodash:BSTcontrol}
\title{Analysis of UAV Communications in \\ Cell-Free Massive MIMO systems}

\author{Carmen D'Andrea, \IEEEmembership{Member, IEEE}, Adrian Garcia-Rodriguez, \IEEEmembership{Member, IEEE}, Giovanni Geraci, \IEEEmembership{Senior Member, IEEE}, \\ Lorenzo Galati Giordano, \IEEEmembership{Member, IEEE}, and Stefano Buzzi, \IEEEmembership{Senior Member, IEEE}
\thanks{Carmen D'Andrea and Stefano Buzzi are with the \textit{Department of Electrical and Information Engineering}, \textit{University of Cassino and Southern Lazio}, I-03043 Cassino, Italy (email:\{buzzi, carmen.dandrea\} @unicas.it) and also with
the \textit{Consorzio Nazionale Interuniversitario per le Telecomunicazioni} (\textit{CNIT}), I-43124 Parma, Italy. Adrian Garcia-Rodriguez and Lorenzo Galati Giordano are with \textit{Nokia Bell Labs}, Dublin, Ireland, \{adrian.garcia\_rodriguez, lorenzo.galati\_giordano\}@nokia-bell-labs.com. Giovanni Geraci is with \textit{Universitat Pompeu Fabra}, Barcelona, Spain, giovanni.geraci@upf.edu. The work of G. Geraci was partly supported by MINECO under Project RTI2018-101040-A-I00 and by the Postdoctoral Junior Leader Fellowship Programme from ``la Caixa'' Banking Foundation.}
}

\IEEEtitleabstractindextext{\begin{abstract}We study support for unmanned aerial vehicle (UAV) communications through a cell-free massive MIMO architecture, wherein a large number of access points (APs) is deployed in place of large co-located massive MIMO arrays. We consider also a variation of the pure cell-free architecture by applying a user-centric association approach, where each user is served only from a subset of APs in the network. Under the general assumption that the propagation channel between the mobile stations, either UAVs or ground users (GUEs), and the APs follows a Ricean distribution, we derive closed form spectral efficiency lower bounds for uplink and downlink with linear minimum mean square error channel estimation. We consider several power allocation and user scheduling strategies for such a system, and, among these, also minimum-rate maximizing power allocation strategies to improve the system fairness. Our numerical results reveal that cell-free massive MIMO architecture and its low-complexity user-centric alternative may provide better performance than a traditional multi-cell massive MIMO network deployment.\end{abstract}


}

\maketitle
\newcounter{MYtempeqncnt}
\section{INTRODUCTION}

\IEEEPARstart{U}nmanned aerial vehicles (UAVs)---also referred to as \textit{drones}---have attracted a great deal of attention in the last few years, both in industry and accademia, due to their ability of performing a wide variety of critical tasks efficiently and in an automated manner. The integration of UAVs in wireless communication networks has thus become a hot research area, mainly with two different approaches \cite{geraci2019preparing,MozSaaBen2019,Mozaffari2016_TWC,Hayat_Survey_2016,Wand_VTM2017,DBLP:journals/corr/abs-1809-01752,vinogradov2019tutorial}. The first research approach focuses on the services that UAVs can provide to wireless networks, since UAVs can be regarded as moving access points (APs). With this perspective, UAVs can be used to increase the network capacity on-demand, fill network coverage holes, fastly deploy a mobile network architecture in the presence of a catastrophic event, etc. \cite{merwaday2015uav,lyu2017placement,mozaffari2016efficient,Bor_Yaliniz_ICC2016}. The second research approach focuses on the communications services that wireless networks can provide to UAVs  \cite{LinYajMur2018,AzaRosPol2017,ZenLyuZha2019,LopDinLi2018GC,azari2019cellular,Mei_TWC_2019}. Considering the latter approach, \cite{GarGerLop2018,GeraciUAVs_Access2018,GerGarGalICC2018,Chandhar_TWC_2018,Amer_CL_UAV2019}
have recently investigated the use of massive MIMO (mMIMO) to support UAVs cellular communications, showing that equipping base stations (BSs) with large antenna arrays dramatically increases---with respect to a traditional cellular deployment---the probability of meeting the stringent reliability requirements of the UAVs command and control (C$\&$C) links.
Additionally, reference \cite{Lyu_Network_Connected_UAV_2019} proposed a new 3D channel model network-connected UAVs and presented a coverage analysis applicable to different network deployments.

In parallel to the research on UAV communications, there has been growing interest about cell-free (CF) mMIMO deployments \cite{Ngo_CellFree2016}, wherein large co-located antenna arrays are substituted by a large number of simpler APs equipped with few antennas and reduced signal processing capabilities. In the CF mMIMO architecture, the APs are connected via a backhaul network to a central processing unit (CPU), which sends to the APs the data symbols to be transmitted to the users and receives soft estimates of the received data symbols from all APs. Neither channel estimates nor beamforming vectors are propagated through the backhaul network, and the time-division-duplex protocol is used to exploit uplink/downlink channel reciprocity. Radio stripes could enable the practical deployment of these systems\cite{Radio_Stripe_patent,interdonato2018ubiquitous}. The results in \cite{Ngo_CellFree2016} show that the CF approach  provides better performance than a small-cell system in terms of 95$\%$-likely per-user throughput. Additionally, \cite{buzziWCL2017, Buzzi_WSA2017, Buzzi_cell_Free_TWC2019} have recently introduced a user-centric (UC) virtual-cell massive MIMO approach to CF mMIMO, assuming that each AP does not serve all the users in the system, but only a subset of them. Overall, the UC approach could be deemed as a low-complexity alternative to CF mMIMO, since APs focus their available resources on the users that will benefit the most from them. CF mMIMO network deployments with UC association rules between the users and the APs are expected to be one of the key technologies for future beyond-5G and 6G wireless networks. Recently, article \cite{Interdonato_ScalabilityICC2019} has discussed some scalability aspects of CF mMIMO systems presenting a scalable implementation of the UC association rule, where one user can be associated to some APs connected to different CPUs. 

\subsection*{Paper contribution}

This work, extending preliminary results reported in the conference paper \cite{DAndrea_UAV_ICC}, analyzes and compares the CF and UC network behaviors in the presence of communication with both UAVs and legacy ground users (GUEs). The paper contributions can be summarized as follows:

\begin{enumerate}

\item Assuming a Ricean channel model for both the GUEs and the UAVs, and linear minimum mean square error (LMMSE) channel estimation, we derive closed-form expressions of the lower-bound of the achievable spectral efficiencies with matched filtering for both the uplink and downlink.
\item We consider and evaluate the performance of several power allocation strategies  for the considered architectures:
\begin{itemize}
\item[-] \emph{For the downlink}, i) we consider a proportional power allocation strategy, ii) we introduce a waterfilling-based power allocation for the CF approach only, and iii) we derive a power control rule aimed at the maximization of the minimum of the spectral efficiencies across the users, using the successive lower-bound maximization technique
 \cite{RazaviyaynSIAM,BuzziZappone_PIMRC2017,AloBuZap5GWF2018}. 
Since experimental evaluation of these power control rules will show that UAVs tend to take a larger share of resource due to their more favorable propagation conditions, these resource allocation rules are also revisited under the constraint that the UAVs and the GUEs  share a pre-determined and fixed set of the system resources. 
\item[-] \emph{For the uplink}, we derive and examine the performance of both i) a fractional power control rule, and ii) a minimum-rate maximizing with resource allocation strategy.
\end{itemize}

\item
To the best of our knowledge, our article performs the first evaluation of CF and UC network deployments with cellular-connected UAVs, and the comparison of these architectures with a customary multi-cell mMIMO network. Our results reveal that CF and UC architectures can outperform multi-cell mMIMO networks for the large majority of UAVs and GUEs in the network as long as 
adequate power and resource allocation procedures are implemented.

\end{enumerate}

The remainder of this paper is organized as follows. Sec.\ \ref{System_model_section} contains the description of the considered system model and the communication process. Sec.\ \ref{Spectral_Efficiencies} includes the derivations of the uplink and downlink spectral efficiency bounds. In Sections \ref{Power_allocation_section_D} and \ref{Power_allocation_section_U} several power allocation strategies are detailed for the downlink and for the uplink, respectively. Sec.\ \ref{Numerical_results} contains the numerical results and presents the key insights of our analysis, while our concluding remarks are given in Sec.\ \ref{conclusions_section}.

\textit{Notation}: In this paper, the following notation is used. $\mathbf{A}$ is a matrix; $\mathbf{a}$ is a vector; $a$ is a scalar. The operators $(\cdot)^T$, $(\cdot)^{-1}$, and $(\cdot)^H$ stand for transpose, inverse and, conjugate transpose, respectively. The determinant of the matrix $\mathbf{A}$ is denoted as $|\mathbf{A}|$ and $\mathbf{I}_P$ is the $P \times P$ identity matrix. The trace of the matrix $\mathbf{A}$ is denoted as tr$\left(\mathbf{A}\right)$. The statistical expectation operator is denoted as $\mathbb{E}[\cdot]$, the real part operator is denoted as $\Re \left\lbrace \cdot \right \rbrace$; $\mathcal{CN}\left(\mu,\sigma^2\right)$ denotes  a complex circularly symmetric Gaussian random variable (RV) with mean $\mu$ and variance $\sigma^2$.

\section{SYSTEM MODEL} \label{System_model_section}
\subsection{Cell-Free Network Topology}

As depicted in Fig. \ref{Fig:Cell_fre_UAV_scenario_Ref_Sys}(a), we consider a network that consists of outdoor APs, GUEs, and UAVs, whose sets are denoted by $\mathcal{A}$, $\mathcal{G}$, and $\mathcal{U}$, and have cardinalities $N_\mathrm{A}$, $N_\mathrm{G}$, and $N_\mathrm{U}$, respectively. In the following, we let the term \emph{users} denote both GUEs and UAVs. The $N_\mathrm{A}$ APs are connected by means of a backhaul network to a CPU wherein data-decoding is performed\footnote{In this article, we assume perfect links between the APs and the CPU, the consideration of limited backhaul links is out of the scope of this paper, an interested reader can be referred to papers \cite{cell_free_limited_backhaul_Burr_ICC2018,Bashar_Burr_TCOM2019}.}. In keeping with the approach of \cite{ngo2015cell,Ngo_CellFree2016}, all communications take place on the same frequency band, i.e. uplink and downlink are separated through time-division-duplex (TDD).

We assume the UAVs and GUEs are equipped with a single antenna, while each AP is equipped with a uniform linear array (ULA) comprised of $N_{\rm AP}$ antennas. We let $\mathcal{K} = \mathcal{G} \cup \mathcal{U}$, and define $K \triangleq N_\mathrm{G}+N_\mathrm{U}$ as the total number of \emph{users} in the system. Let us also denote by $\mathcal{K}_a$ the set of users served by the $a$-th AP on a given physical resource block (PRB), and by $K_a$ its cardinality. The set of APs serving user $k$ is denoted by $\mathcal{A}_k$. The set of users associated to each AP can be determined according to several criteria. In general, the user association can be formulated as an integer optimization problem, whose solution is not straightforward. In this paper, we consider the two following a-priori approaches.

\subsubsection{CF approach}
In the CF approach, each AP communicates with all the users in the system, i.e.,  we have that $\mathcal{K}_a = \mathcal{K}, \; \forall \, a=1,\ldots,N_\mathrm{A}$ and the set $\mathcal{A}_k=\mathcal{A}, \; \forall \, k=1,\ldots,K$.

\subsubsection{UC approach}
In the UC approach, the $k$-th user is served by the $A_k$ APs that it receives with best average channel conditions. Let $\beta_{k,a}$ characterize the scalar coefficient modeling the channel path-loss and shadowing effects between the $k$-th user and the $a$-th AP and $O_k \, : \, \{1,\ldots, N_\mathrm{A} \} \rightarrow \{1,\ldots, N_\mathrm{A} \}$ denote the sorting operator for the vector $\left[\beta_{k,1},\ldots, \beta_{k,N_\mathrm{A}}\right]$, such that $\beta_{k,O_k(1)} \geq \beta_{k,O_k(2)} \geq \ldots \geq \beta_{k,O_k(N_\mathrm{A})}$. The set $\mathcal{A}_k$ of the $A_k$ APs serving the $k$-th user is then given by
\begin{equation}
\mathcal{A}_k=\{ O_k(1), O_k(2), \ldots , O_k(A_k) \}.
\end{equation}
Consequently, the set of users served by the $a$-th AP is defined as  $\mathcal{K}_a=\{ k: \,  a \in \mathcal{A}_k \}$.

\begin{figure*}[t!]
	\centering
    \begin{subfigure}[b]{0.5\textwidth}
    \centering
	\includegraphics[scale=0.4]{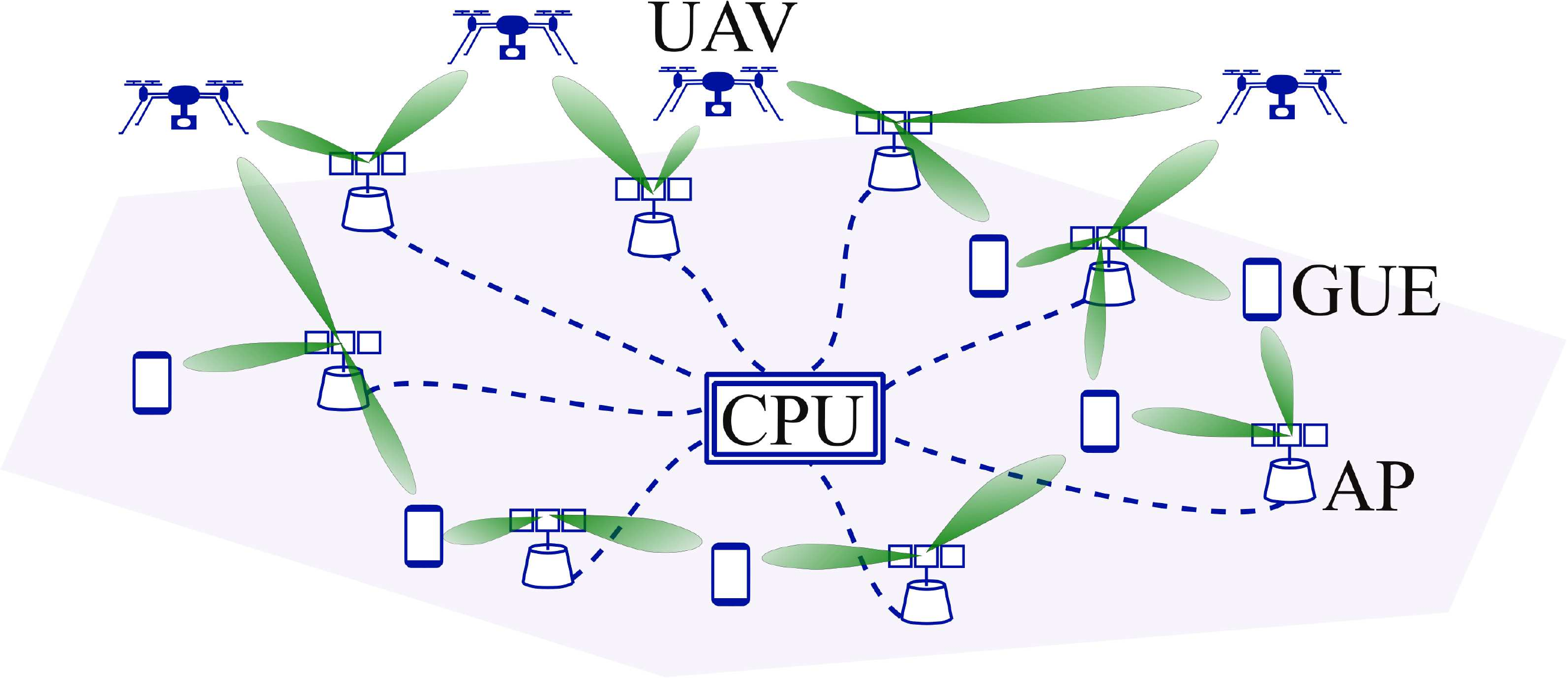}
        \caption{\color{white}.}
        \label{Fig:Cell_fre_UAV_scenario_Ref_Sys_a}
    \end{subfigure}%
    \begin{subfigure}[b]{0.5\textwidth}
       \centering
\includegraphics[scale=0.5]{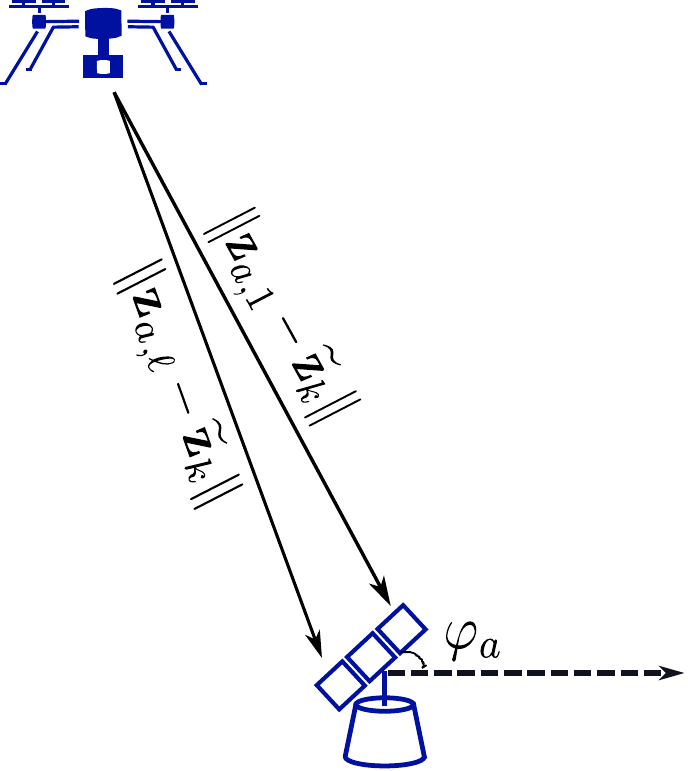}
        \caption{\color{white}.}
        \label{Fig:Cell_fre_UAV_scenario_Ref_Sys_b}
    \end{subfigure}
    \caption{In (a) a CF network supporting both ground and UAV users, and in (b) the reference system for the phase rotation evaluation in the channel between a generic AP-UAV pair.}
    \label{Fig:Cell_fre_UAV_scenario_Ref_Sys}
\end{figure*}

\subsection{Propagation Channel}

We denote by $\mathbf{g}_{k,a} \in \mathbb{C}^{N_{\rm AP}}$ the channel between the $k$-th user and the $a$-th AP. Throughout the paper we characterize the small-scale fading through a Ricean fading model, which consists of a dominant line-of-sight (LOS) component and a Rayleigh–distributed factor modelling the rich-scattered multipath. The channel between the $k$-th generic user and the $a$-th AP is thus written as

\begin{equation}
\mathbf{g}_{k,a}=\sqrt{\frac{\beta_{k,a}}{K_{k,a}+1}} \left[ \sqrt{K_{k,a}} e^{j \vartheta_{k,a}} \mathbf{a}\left(\theta_{k,a}\right) + \mathbf{h}_{k,a}  \right] \; ,
\label{Channel_generic}
\end{equation}
where $K_{k,a}$ is the Ricean $K$-factor, $\mathbf{h}_{k,a} \in \mathbb{C}^{N_{\rm AP}}$ contains the small-scale fading i.i.d. $\mathcal{CN} (0,1)$ coefficients between the $a$-th AP and the $k$-th user, and $\vartheta_{k,a}$ follows a uniform distribution in $[0, 2\pi]$, denoting the random phase offset for the direct path. Moreover, $\mathbf{a}\left(\theta_{k,a}\right) \in \mathbb{C}^{N_{\rm AP}}$ represents the steering vector evaluated at the angle $\theta_{k,a}$, which characterizes the direct path between the $a$-th AP and the $k$-th user. As illustrated in Fig. \ref{Fig:Cell_fre_UAV_scenario_Ref_Sys_b}, denoting by $d$ the antenna spacing at the AP, and by $\mathbf{z}_{a,q}$ and $\widetilde{\mathbf{z}}_k$ the 3D vectors containing the position of the $q$-th antenna element at the $a$-th AP and of the $k$-th user, respectively, 
the $\ell$-th entry of the vector $\mathbf{a}(\theta_{k,a})$ can be written as
$
\left[\mathbf{a}\left(\theta_{k,a}\right)\right]_{\ell}=e^{-j\frac{2\pi}{\lambda}\left( \|\mathbf{z}_{a,1}-\widetilde{\mathbf{z}}_k \| - \|\mathbf{z}_{a,\ell}-\widetilde{\mathbf{z}}_k \|\right)},
$ with $\ell=1,\ldots,N_{\rm AP}$.

We consider that the Ricean $K$-factor $K_{k,a}$ depends on the probability that the link between the $k$-th user and the $a$-th AP $p_{\rm LOS}\left(d_{k,a}\right)$ is LOS, which, in turn, depends on the link length $d_{k,a}$  following \cite{Jafari_Lopez_2015}. We thus have
$K_{k,a}=\frac{p_{\rm LOS}\left(d_{k,a}\right)}{1-p_{\rm LOS}\left(d_{k,a}\right)}
$.

\subsection{The Communication Process: Uplink Training}
\label{MMSE_Ch_est}

Let us denote by $\tau_c$ the dimension in time/frequency samples of the channel coherence length, and by $\tau_p < \tau_c$ the dimension of the uplink training phase. 
We also define $\boldsymbol{\phi}_k \in \mathbb{C}^{\tau_p}$ as the pilot sequence sent by user $k$, and assume that $\|\boldsymbol{\phi}_k\|^2=1 \forall k$.
The signal received at the $a$-th AP during the training phase $\mathbf{Y}_a \in \mathbb{C}^{N_{\rm AP}  \times \tau_p}$ can be therefore expressed as
\begin{equation}
\mathbf{Y}_a = \ds \sum_{k \in \mathcal{K}} \ds \sqrt{\eta_k} \mathbf{g}_{k,a}\boldsymbol{\phi}_k^H + \mathbf{W}_a \; ,
\label{eq:y_m}
\end{equation}
where ${\eta}_k$ denotes the power employed by the $k$-th user during the training phase, and $\mathbf{W}_a \in \mathbb{C}^{N_{\rm AP}  \times \tau_p}$ contains the thermal noise contribution and out-of-cell interference at the $a$-th AP, with i.i.d. ${\cal CN}(0, \sigma^2_w)$ RVs as entries.

From the observable $\mathbf{Y}_a$, and exploiting the knowledge of the users' pilot sequences, the $a$-th AP can estimate the channel vectors 
$\left\{\mathbf{g}_{k,a}\right\}_{k\in \mathcal{K}_a}$ based on the statistics
\begin{equation}
\widehat{\mathbf{y}}_{k,a}=\mathbf{Y}_a \boldsymbol{\phi}_k= \sqrt{\eta_k}\mathbf{g}_{k,a} + \ds \sum_{\substack{i=1 \\ i\neq k}}^K {\sqrt{\eta_i}\mathbf{g}_{i,a}\boldsymbol{\phi}_i^H \boldsymbol{\phi}_k} + \mathbf{W}_a \boldsymbol{\phi}_k \; .
\label{y_hat_ka}
\end{equation}
Specifically, assuming knowledge of the large-scale fading coefficients $\beta_{k,a}$ as in \cite{Ngo_CellFree2016}, and of the vectors $\mathbf{a}\left(\theta_{k,a}\right) \, \forall \; a,k$, the LMMSE estimate of $\left\{\mathbf{g}_{k,a}\right\}_{k\in \mathcal{K}_a}$ of the channel $\mathbf{g}_{k,a}$ is given by\footnote{Note that we are assuming at the APs the knowledge of the vectors $\mathbf{a}\left(\theta_{k,a}\right)$, i.e., of the directions of arrival of the main LoS paths. This assumption is realistic because the main direction of arrival can be easily estimated at the APs and it changes approximately on the time-scale of the variation of the path-loss and shadow fading coefficients $\beta_{k,a}$.}

\begin{equation}
\widehat{\mathbf{g}}_{k,a}= \mathbf{D}_{k,a} \,  \widehat{\mathbf{y}}_{k,a} \, ,
\end{equation}

where 

\begin{equation*}
\begin{array}{lllll}
\mathbf{D}_{k,a}= \sqrt{\eta_k} \mathbf{G}_{k,a} \mathbf{B}_{k,a}^{-1} \in \mathbb{C}^{N_{\rm AP} \times N_{\rm AP}} \, , \\
\mathbf{G}_{k,a} = \frac{\beta_{k,a}}{K_{k,a}+1} \left[ K_{k,a} \mathbf{a}\left(\theta_{k,a}\right)\mathbf{a}^H\left(\theta_{k,a}\right) +\mathbf{I}_{N_{\rm AP}}\right] \, , \\
\mathbf{B}_{k,a} =\sum_{i \in \mathcal{K}} {\eta_i \beta_{i,a} \mathbf{G}_{i,a}\left|\boldsymbol{\phi}_i^H \boldsymbol{\phi}_k\right|^2 } + \sigma^2_w \mathbf{I}_{N_{\rm AP}}
\end{array}
\end{equation*}

\subsection{The Communication Process: Downlink Data Transmission} \label{DL_Section}

The APs treat the channel estimates as the true channels and perform conjugate beamforming on the downlink. The signal transmitted by the $a$-th AP in a generic symbol interval is the following $N_{\rm AP}$-dimensional vector
\begin{equation}
\mathbf{s}_a = \ds \sum_{k\in{\cal K}_a}\ds \sqrt{\eta_{k,a}^{\rm DL}} \widehat{\mathbf{g}}_{k,a} {x}_k^{\rm DL} \; ,
\label{eq:transmittedscalar}
\end{equation}
with 
${x}_k^{\rm DL}$ being the downlink data-symbol for the $k$-th user, and $\eta_{k,a}^{\rm DL}$ a scalar coefficient controlling the power transmitted by the $a$-th AP to the $k$-th user. Letting $\eta^{\rm DL}_a$ denote the overall transmitted power by the $a$-th AP, the normalized transmit power must satisfy the constraint
\begin{equation}
\mathbb{E} \left[ \|\mathbf{s}_a \|^2\right]=\ds \sum_{k\in{\cal K}_a} {\eta_{k,a}^{\rm DL} \gamma_{k,a}} \leq \eta^{\rm DL}_a \; ,
\end{equation}
where 
$
\gamma_{k,a}=\mathbb{E}\left[ \widehat{\mathbf{g}}_{k,a}^H \widehat{\mathbf{g}}_{k,a}\right]=\sqrt{\eta_k}\text{tr} \left( \mathbf{G}_{k,a} \mathbf{D}_{k,a} \right).
$

Subsequently, each user receives phase-aligned contributions from all APs. In particular, the $k$-th user receives the soft estimate for the data symbol
\begin{equation}
\begin{array}{llll}
\widehat{x}_k^{\rm DL} = & \ds \sum_{a \in \mathcal{A}}
\mathbf{g}_{k,a}^H \mathbf{s}_a + {z}_k  
 = \ds \sum_{a\in{\cal A}_k} \ds \sqrt{\eta_{k,a}^{\rm DL}} \mathbf{g}_{k,a}^H  \widehat{\mathbf{g}}_{k,a} {x}_k^{\rm DL}  \\ & + 
\ds \sum_{j \in \mathcal{K}\backslash k} \ds \sum_{a \in \mathcal{A}_j} \sqrt{\eta_{j,a}^{\rm DL}}\mathbf{g}_{k,a}^H   \widehat{\mathbf{g}}_{j,a} {x}_j^{\rm DL} +  
 {z}_k\; ,
 \end{array}
\label{eq:received_data_MS_UC}
\end{equation}
with ${z}_k$ being the ${\cal CN}(0, \sigma^2_z)$ additive white Gaussian noise (AWGN).

\subsection{The Communication Process: Uplink Data Transmission} 
\label{UL_Section}

In uplink, users send their data symbols without any channel-dependent phase offset. As a result, the signal $\bar{\mathbf{y}}_a \in \mathbb{C}^{N_{\rm AP}}$ received at the $a$-th AP in a generic symbol interval can be expressed as 
\begin{equation}
{\bar{\mathbf{y}}}_a=\ds \sum_{k \in \mathcal{K}} \ds \sqrt{\eta_{k}^{\rm UL}} \mathbf{g}_{k,a} {x}^{\rm UL}_k + \mathbf{w}_a \; ,
\end{equation}
with ${\eta_{k}^{\rm UL}}$ and ${x}^{\rm UL}_k$  representing the uplink transmit power and the data symbol of the $k$-th user, respectively, and $\mathbf{w}_a \sim {\cal CN}(\mathbf{0}, \sigma^2_w \mathbf{I} ) \in \mathbb{C}^{N_{\rm AP}}$ the AWGN vector.

Subsequently, each AP decodes the data transmitted by users in ${\cal K}_a$. The $a$-th AP thus forms, for each $k \in {\cal K}_a$,  the  statistics
${{t}}_{a,k}= \widehat{\mathbf{g}}_{k,a}^H {\bar{\mathbf{y}}}_a$ and sends them to the CPU.  Accordingly, the CPU is able to derive soft estimates
\begin{equation}
\widehat{{x}}^{\rm UL}_k = \ds \sum_{a \in \mathcal{A}_k} {{t}}_{a,k} \; , \quad k \in \mathcal{K} \, 
\label{Est_UL_uc1_MS}
\end{equation}
of the data sent by the users.

\section{SPECTRAL EFFICIENCY BOUNDS} \label{Spectral_Efficiencies}
In this section, we report lower and upper spectral efficiency bounds both for the downlink and uplink data transmission phases.

\subsection{Downlink Data Transmission} 
\label{DL_Bounds}

\subsubsection{Lower Bound}
\label{subsec:DLLowerBound}

A LB for the downlink spectral efficiency can be also computed, based on the assumption that each user has knowledge of the channel statistics but not of the channel realizations. The received signal in \eqref{eq:received_data_MS_UC} can be thus rewritten as

\begin{equation}
\begin{array}{llll}
&\widehat{x}_k^{\rm DL} =    \underbrace{\mathbb{E} \left[\ds \sum_{a\in{\cal A}_k} \ds \sqrt{\eta_{k,a}^{\rm DL}} \mathbf{g}_{k,a}^H  \widehat{\mathbf{g}}_{k,a} \right]}_{D_k} {x}_k^{\rm DL} \\ & +
\underbrace{\left( \ds \sum_{a\in{\cal A}_k} \ds \sqrt{\eta_{k,a}^{\rm DL}} \mathbf{g}_{k,a}^H  \widehat{\mathbf{g}}_{k,a} - \mathbb{E} \left[\ds \sum_{a\in{\cal A}_k} \ds \sqrt{\eta_{k,a}^{\rm DL}} \mathbf{g}_{k,a}^H  \widehat{\mathbf{g}}_{k,a} \right] \right)}_{B_k}{x}_k^{\rm DL} \\ & +
\ds \sum_{j \in \mathcal{K}\backslash k} \ds \underbrace{\sum_{a \in \mathcal{A}_j} \sqrt{\eta_{j,a}^{\rm DL}}\mathbf{g}_{k,a}^H   \widehat{\mathbf{g}}_{j,a}}_{I_{k,j}} {x}_j^{\rm DL} +  
 {z}_k\; ,
\end{array}
\label{eq:received_data_MS_UC2}
\end{equation}
where $D_k$, $B_k$, and $I_{k,j}$ represent the strength of desired signal, the beamforming gain uncertainty, and the interference caused by
the $k$-th user, respectively.
We treat the sum of the second, third, and fourth summand  in \eqref{eq:received_data_MS_UC2} as ``effective noise'' as in references \cite{Ngo_CellFree2016,CF_Rician_Bjornson_2018,Ngo_CF_Ricean2018}.
By using the fact that uncorrelated Gaussian noise represents the worst case, we obtain the following LB for the  downlink spectral efficiency of the $k$-th user in the system:

\begin{equation}
\text{SE}_{k, {\rm LB}}^{\rm DL}=\frac{\tau_{\rm d}}{\tau_c} \log_2 \left(1 + \frac{\left|D_{k}\right|^2}{\mathbb{E}\left[ \left|B_{k}\right|^2\right]+\ds \sum_{j \in \mathcal{K}\backslash k} {\mathbb{E}\left[ \left|I_{k,j}\right|^2\right]} + \sigma^2_z}\right),
\label{eq:SE_general}
\end{equation}
where $\tau_{\rm d}= \tau_c - \tau_p -\tau_{\rm u}$ and $\tau_{\rm u} $ are the lengths (in time-frequency samples) of  the downlink and uplink data transmission phases in each coherence interval, respectively.

Eq. \eqref{eq:SE_general} is deterministic and contains several expectations over the random channel realizations.
For the general case, these expectations are not available in closed form but can be computed through Monte Carlo simulations. Conversely, for the case of conjugate beamforming and LMMSE channel estimation, a closed form expression for the  spectral efficiency LB can be obtained. We have indeed the following result:

\begin{figure*}[t!]
\begin{equation}
\begin{aligned}
&\!\!\!\!\!\!\!\!\!\overline{\text{SINR}}_{k, {\rm DL}} =\!\! \ds \left( \ds \sum_{a\in{\cal A}_k} {\ds \sqrt{\eta_{k,a}^{\rm DL}} \gamma_{k,a}} \right)^2 \!\!\!\! \times \!\!\Bigg\{\!\!
\ds \sum_{a\in{\cal A}_k} \eta_{k,a}^{\rm DL} \left( \eta_{k} \delta_{k,a}^{(k)}-\gamma_{k,a}^2\right)\!\!+ \!\!\ds \sum_{j \in \mathcal{K}} \sqrt{\eta_j} \ds \sum_{a\in{\cal A}_j} \eta_{j,a}^{\rm DL} \text{tr} \left( \mathbf{G}_{j,a} \mathbf{D}_{j,a}^H \mathbf{G}_{k,a} \right) \\ + \sigma^2_z 
& \enspace + \ds \sum_{j \in \mathcal{K}\backslash k} \eta_k \bigg\{ \ds \sum_{a\in{\cal A}_j} \bigg[\eta_{j,a}^{\rm DL}\delta_{k,a}^{(j)} + \ds \sum_{\substack{b\in{\cal A}_j \\ b \neq a}} \sqrt{\eta_{j,a}^{\rm DL}} \sqrt{\eta_{j,b}^{\rm DL}} \text{tr} \left(\mathbf{D}_{j,a}\mathbf{G}_{k,a}\right)\text{tr} \left(\mathbf{D}_{j,b}^H\mathbf{G}_{k,b}\right)\bigg]\bigg\} \left|\boldsymbol{\phi}_k^H \boldsymbol{\phi}_j \right|^2 \Bigg\}^{-1}
\end{aligned}
\label{eq:SINR_bar_DL}
\end{equation}
\hrulefill
\end{figure*}

\emph{Lemma 1:} A LB for the downlink spectral efficiency in the case of conjugate beamforming and LMMSE channel estimation is given by
\begin{equation}
\text{SE}_{k,{\rm LB}}^{\rm DL}= \ds\frac{\tau_{\rm d}}{\tau_c}  \log_2 \left( \ds 1 + \overline{\text{SINR}}_{k, {\rm DL}}\right) \; ,
\label{eq:SE_MMSE_LB}
\end{equation}
where $\overline{\text{SINR}}_{k, {\rm DL}}$ is shown in \eqref{eq:SINR_bar_DL} at the top of the next page, and 
%

\begin{equation}
\begin{array}{llll}
&\delta_{k,a}^{(j)}=\ds \left( \frac{\beta_{k,a}}{K_{k,a}+1}\right)^2 \text{tr}^2\left(\mathbf{D}_{j,a}\right) \\ & + 2 K_{k,a} \left( \ds \frac{\beta_{k,a}}{K_{k,a}+1}\right)^2  \ds \Re \left\lbrace \text{tr} \left( \mathbf{a}^H\left(\theta_{k,a}\right) \mathbf{D}_{j,a} \mathbf{a}\left(\theta_{k,a}\right) \mathbf{D}_{j,a}^H \right) \right\rbrace.
\end{array}
\label{delta_ka_j}
\end{equation}

\emph{Proof: } The proof of Lemma 1 is based on the application of the use-and-then-forget (UatF) bound \cite{marzetta2016fundamentals}. The complete details are reported in Appendix \ref{Appendix_DL}. \hfill $\square$

\subsubsection{Upper Bound}

To provide an intuition about the tightness of the LB derived in Sec.\ \ref{subsec:DLLowerBound}, we also include a UB of the achievable downlink spectral efficiency here. Specifically, given the expression in \eqref{eq:received_data_MS_UC} an upper bound (UB) for the achievable spectral efficiency can be obtained as \cite{Caire_bounds_2018}
\begin{equation}
\text{SE}_{k, {\rm UB}}^{\rm DL}\!=\!\ds\frac{\tau_{\rm d}}{\tau_c}\mathbb{E}\!\! \left[ \! 1+ \frac{\ds \left|\sum_{a\in{\cal A}_k}  \sqrt{\eta_{k,a}^{\rm DL}} \mathbf{g}_{k,a}^H  \widehat{\mathbf{g}}_{k,a}\right|^2}{\ds \sum_{j \in \mathcal{K}\backslash k} \left| \ds \sum_{a \in \mathcal{A}_j} \sqrt{\eta_{j,a}^{\rm DL}}\mathbf{g}_{k,a}^H   \widehat{\mathbf{g}}_{j,a}\right|^2\!\!\!\!+\!\sigma^2_z} \!\! \right] \, .
\label{SE_DL_upper_bound}
\end{equation}
The expectation in \eqref{SE_DL_upper_bound} is made over the fast fading channel realizations. 

\subsection{Uplink Data Transmission} 
\label{UL_Bounds}

\subsubsection{Lower Bound}
\label{subsec:ULLowerBound}

Using straightforward manipulations, \eqref{Est_UL_uc1_MS} can be re-written as
\begin{equation}
\begin{array}{llll}
\widehat{x}_k^{\rm UL} = & \ds \sum_{a\in{\cal A}_k} \ds \sqrt{\eta_{k}^{\rm UL}} \widehat{\mathbf{g}}_{k,a}^H  \mathbf{g}_{k,a} {x}_k^{\rm UL} \\ & +
\ds \sum_{j \in \mathcal{K}\backslash k} \ds \sum_{a \in \mathcal{A}_k} \sqrt{\eta_{j}^{\rm UL}}\widehat{\mathbf{g}}_{k,a}^H   \mathbf{g}_{j,a} {x}_j^{\rm UL} +  
 \ds \sum_{a\in{\cal A}_k} {\widehat{\mathbf{g}}_{k,a}^H \mathbf{w}_a  } .
 \end{array}
\label{Est_UL_uc1_MS2}
\end{equation}

In order to derive a LB for the uplink spectral efficiency,  we assume that the CPU 
relies only on statistical knowledge of the channel coeficients when performing the detection, so that \eqref{Est_UL_uc1_MS2} can be re-written as
\begin{equation}
\begin{array}{llll}
&\widehat{x}_k^{\rm UL} = \underbrace{ \mathbb{E} \left[\ds \sum_{a\in{\cal A}_k} \ds \sqrt{\eta_{k}^{\rm UL}} \widehat{\mathbf{g}}_{k,a}^H  \mathbf{g}_{k,a}\right]}_{\widetilde{D}_k} {x}_k^{\rm UL} \\ & + \underbrace{\left( \ds \sum_{a\in{\cal A}_k} \ds \sqrt{\eta_{k}^{\rm UL}} \widehat{\mathbf{g}}_{k,a}^H  \mathbf{g}_{k,a} - \mathbb{E} \left[\ds \sum_{a\in{\cal A}_k} \ds \sqrt{\eta_{k}^{\rm UL}} \widehat{\mathbf{g}}_{k,a}^H  \mathbf{g}_{k,a}\right] \right)}_{\widetilde{B}_k} {x}_k^{\rm UL} \\ & +
\ds \sum_{j \in \mathcal{K}\backslash k}  \underbrace{\ds \sum_{a \in \mathcal{A}_k} \sqrt{\eta_{j}^{\rm UL}}\widehat{\mathbf{g}}_{k,a}^H   \mathbf{g}_{j,a}}_{\widetilde{I}_{k,j}} {x}_j^{\rm UL} +  
 \underbrace{\ds \sum_{a\in{\cal A}_k} {\widehat{\mathbf{g}}_{k,a}^H \mathbf{w}_a}}_{\widetilde{N}_k} \; .
\end{array}
\label{Est_UL_uc1_MS3}
\end{equation}
Again we model the sum of the second, third, and fourth summand as ``effective noise'' and use the worst-case gaussian assumption, which leads to the following UB for the uplink spectral efficiency of the $k$-th user:
\begin{equation}
\!\!\text{SE}_{k, {\rm LB}}^{\rm UL}\!\!= \!\!\frac{\tau_{\rm u}}{\tau_c} \log_2 \!\!\left( \!1 \! + \!\!  \ds \frac{\left|\widetilde{D}_{k}\right|^2}{\!\mathbb{E}\left[ \left|\widetilde{B}_{k}\right|^2\right]\!\!\!+\!\!\!\!\!\ds \sum_{j \in \mathcal{K}\backslash k} {\mathbb{E}\left[ \left|\widetilde{I}_{k,j}\right|^2\right]} \!\!\!+ \!\!\mathbb{E}\left[ \left|\widetilde{N}_{k}\right|^2\right] }\right).
\label{eq:SE_general_UL}
\end{equation} 
  
As for the downlink, Eq. \eqref{eq:SE_general_UL} is deterministic and contains several expectations over the random channel realizations that can be computed, in the general case, by means of Monte Carlo simulations. Conversely, for the case in which  matched filter detection  and LMMSE channel estimation are used, a closed-form expression can be worked out. We have indeed the following result:

\begin{figure*}[b!]
\hrulefill
\begin{equation}
\begin{aligned}
\!\!&\overline{\text{SINR}}_{k, {\rm UL}}\!\! = \ds \eta_{k}^{\rm UL} \!\! \left( \ds \sum_{a\in{\cal A}_k} {\ds  \gamma_{k,a}} \right)^2 \!\!\times \!\!\Bigg\{ \ds \eta_{k}^{\rm UL}\!\!\! \sum_{a\in{\cal A}_k}  \left( \eta_{k} \widetilde{\delta}_{k,a}^{(k)}-\gamma_{k,a}^2\right)\!\!+ \!\!\ds \sum_{j \in \mathcal{K}} \eta_j^{\rm UL} \sqrt{\eta_k} \ds \sum_{a\in{\cal A}_k}  \text{tr} \left( \mathbf{G}_{k,a} \mathbf{D}_{k,a}^H \mathbf{G}_{j,a} \right)\\
 & \enspace + \sigma^2_w \sum_{a\in{\cal A}_k} {\ds  \gamma_{k,a}} + \ds \sum_{j \in \mathcal{K}\backslash k} \eta_j^{\rm UL} \eta_j \bigg\{ \ds \sum_{a\in{\cal A}_k} \bigg[\widetilde{\delta}_{j,a}^{(k)} + \ds \sum_{\substack{b\in{\cal A}_k \\ b \neq a}}  \text{tr} \left(\mathbf{D}_{k,a}^H\mathbf{G}_{j,a}\right)\text{tr} \left(\mathbf{D}_{k,b}\mathbf{G}_{j,b}\right)\bigg]\bigg\} \left|\boldsymbol{\phi}_j^H \boldsymbol{\phi}_k \right|^2 \Bigg\}^{-1}.
\end{aligned}
\label{eq:SINR_bar_UL}
\end{equation}
\end{figure*}

\emph{Lemma 2: } Assuming that data decoding is performed through matched filtering and LMMSE channel estimation is used, a LB for the $k$-th user uplink spectral efficiency  can be expressed as
\begin{equation}
\text{SE}_{k,{\rm LB}}^{\rm UL}= \ds\frac{\tau_{\rm u}}{\tau_c}   \log_2 \left( \ds 1 + \overline{\text{SINR}}_{k, {\rm UL}}\right) \; ,
\label{eq:SE_MMSE_LB_UL}
\end{equation}
where $\overline{\text{SINR}}_{k, {\rm UL}}$ is reported in \eqref{eq:SINR_bar_UL} at the bottom of the next page, and


\begin{equation}
\begin{array}{llll}
&\widetilde{\delta}_{j,a}^{(k)}=\ds \left( \frac{\beta_{j,a}}{K_{j,a}+1}\right)^2 \text{tr}^2\left(\mathbf{D}_{k,a}\right) + \\ & 2 K_{j,a} \ds \left( \frac{\beta_{j,a}}{K_{j,a}+1}\right)^2 \Re \left\lbrace \text{tr} \left( \mathbf{a}^H\left(\theta_{j,a}\right) \mathbf{D}_{k,a} \mathbf{a}\left(\theta_{j,a}\right) \mathbf{D}_{k,a}^H \right)\right\rbrace.
\end{array}
\label{delta_tilde_ka_j}
\end{equation}

\emph{Proof:}  The proof of Lemma 2 is also based on the application of the UatF bound \cite{marzetta2016fundamentals}. The details of the proof are reported in Appendix \ref{Appendix_UL}.  \hfill $\square$ 

\subsubsection{Upper Bound}

Following steps similar to the ones detailed above for the downlink, a UB for the achievable spectral efficiency can be obtained as \cite{Caire_bounds_2018}

\begin{equation}
\!\text{SE}_{k, {\rm UB}}^{\rm UL}\!=\!\ds\frac{\tau_{\rm u}}{\tau_c} \!  \mathbb{E} \!\!\left[ \! 1\!+\! \frac{\ds \eta_{k}^{\rm UL}\left|\sum_{a\in{\cal A}_k} \widehat{\mathbf{g}}_{k,a}^H  \mathbf{g}_{k,a}\right|^2}{\!\!\!\ds \sum_{j \in \mathcal{K}\backslash k} \!\eta_{j}^{\rm UL}\! \left| \ds \sum_{a \in \mathcal{A}_k} \widehat{\mathbf{g}}_{k,a}^H \mathbf{g}_{j,a}\right|^2\!\!\!+\!\sigma^2_w \!\!\! \sum_{a \in \mathcal{A}_k} \!\!\| \widehat{\mathbf{g}}_{k,a} \|^2}  \right].
\label{SE_UL_upper_bound}
\end{equation}

\section{POWER ALLOCATION STRATEGIES}
In this section we derive and describe a variety of power allocation strategies tailored for networks with coexisting UAVs and GUEs.

\subsection{Downlink Data Transmission}
\label{Power_allocation_section_D}

\subsubsection{Minimum-rate Maximization} \label{Max_Min_PA_DL}

Focusing on downlink data transmission, we start by considering the case in which power allocation is aimed at maximizing the minimum spectral efficiency across users. We will work here with the UB expression for the spectral efficiency as reported in  \eqref{SE_DL_upper_bound}. 

The minimum-rate maximization problem is formulated as
\begin{subequations}\label{Prob:MinRate}
\begin{align}
&\ds\max_{\bbeta}\;\min_{k\in{\cal K}}\; \mathcal{R}_{k}^{\rm DL}(\bbeta)\label{Prob:aMinRate}\\
&\;\textrm{s.t.}\; \sum_{k\in{\cal K}_a}\eta_{k,a}^{\rm DL} \gamma_{k,a} \leq \eta^{\rm DL}_a\;,\forall\;a=1,\ldots,N_A\label{Prob:bMinRate}\\
&\;\;\;\quad\eta_{k,a}^{\rm DL}\geq 0\;,\forall\;a=1,\ldots,N_A,\;k=1,\ldots,K,\label{Prob:cMinRate}
\end{align}
\end{subequations}
where $\mathcal{R}_{k}^{\rm DL}(\bbeta)= W \text{SE}_{k, {\rm UB}}^{\rm DL}$ is the achievable downlink rate, $W$ is the system bandwidth, and $\bbeta$ is the $KM\times 1$ vector collecting the downlink transmit powers of all APs for all users. 

In order to obtain a more tractable form, we reformulate the optimization problem as follows
 \begin{subequations}\label{Prob:MinRate2}
\begin{align}
&\ds\max_{\overline{\bbeta}}\;\min_{k\in{\cal K}}\; \mathcal{R}_{k}^{\rm DL}(\overline{\bbeta})\label{Prob:aMinRate2}\\
&\;\textrm{s.t.}\; \sum_{k\in{\cal K}_a}\overline{\eta}_{k,a}^{\rm DL} \rho_{a,k}^{\rm DL} \gamma_{k,a} \leq \eta^{\rm DL}_a\;,\forall\;a=1,\ldots,N_A\label{Prob:bMinRate2}\\
&\;\;\;\quad \overline{\eta}_{k,a}^{\rm DL}\geq 0\;,\forall\;a=1,\ldots,N_A,\;k=1,\ldots,K,\label{Prob:cMinRate2}
\end{align}
\end{subequations}
where 
$
\rho_{a,k}^{\rm DL}= \left(\ds \sum_{j\in{\cal K}_a} \gamma_{j,a}\right)^{-1},
$
and $ 0 \leq \overline{\eta}_{k,a}^{\rm DL}\leq \eta^{\rm DL}_a\;,\forall\;a=1,\ldots,N_A,\;k=1,\ldots,K$ are the \textit{normalized} transmit powers.

The Problem \eqref{Prob:MinRate} has a non-concave and non-differentiable objective function, and therefore cannot be solved through efficient numerical methods. The following Lemma describes how the problem can be solved in a tractable manner.

\emph{Lemma 3: } The optimization problem \eqref{Prob:MinRate} can be solved using the procedure stated in Algorithm \ref{Alg:MRMax}, where $\mathcal{P}_p$ is the optimization problem
\begin{subequations}\label{Prob:MinRateApp}
\begin{align}
\mathcal{P}_p: &\ds\max_{\overline{\bbeta}_{a}^{(p)},t}\; t \label{Prob:aMinRateApp}\\
&\;\textrm{s.t.}\;\sum_{k\in{\cal K}_a}\overline{\eta}_{k,a}^{\rm DL} \rho_{a,k}^{\rm DL} \gamma_{k,a} \leq \eta^{\rm DL}_a\;,\forall\;a=1,\ldots,N_A\label{Prob:bMinRateApp}\\
&\;\;\;\quad \overline{\eta}_{k,a}^{\rm DL}\geq 0\;,\forall\;a=1,\ldots,N_A,\;k=1,\ldots,K,\label{Prob:cMinRateApp}\\
&\;\;\;\quad \widetilde{\mathcal{R}}_{k}^{\rm DL}\left(\overline{\bbeta}_{a}^{(p)},\overline{\bbeta}_{a,0}^{(p)},\overline{\bbeta}_{-a}^{(-p)}\right) \geq t,\; \forall \; k=1,\ldots,K, \label{Prob:dMinRateApp}
\end{align}
\end{subequations}
which can be easily shown to be convex for any $\overline{\bbeta}_{a,0}^{(p)}$, and therefore can be solved through standard techniques. 

\emph{Proof:} The proof of Lemma 3 is based on the application of the successive lower-bound maximization framework and successive reformulations. The complete details and the definition of $\widetilde{\mathcal{R}}_{k}^{\rm DL}\left(\overline{\bbeta}_{a}^{(p)},\overline{\bbeta}_{a,0}^{(p)},\overline{\bbeta}_{-a}^{(-p)}\right)$ are reported in Appendix \ref{Appendix_Lemma3}. \hfill $\square$

\begin{algorithm}[t]
\caption{Minimum-rate maximization in downlink}
\begin{algorithmic}[1]
\label{Alg:MRMax}
\STATE Set $i=0$ and choose any feasible $\overline{\bbeta}_{2}^{(1)},\ldots, \overline{\bbeta}_{2}^{(P_2)}\ldots,\overline{\bbeta}_{N_A}^{(1)},\ldots,\overline{\bbeta}_{N_A}^{(P_{N_A})}$;
\REPEAT
	\FOR{$a=1\to N_A$}
		\FOR{$p=1\to P_a$} 
		\REPEAT
		\STATE Choose any feasible $\overline{\bbeta}_{a,0}^{(p)}$;
		\STATE Let $\overline{\bbeta}_{a}^{(p),*}$ be the solution of $\mathcal{P}_p$ in \eqref{Prob:MinRateApp};
		\STATE $\overline{\bbeta}_{a,0}^{(p)}=\overline{\bbeta}_{a}^{(p),*}$;
		\UNTIL{convergence}
		\STATE $\overline{\bbeta}_{a}^{(p)}=\overline{\bbeta}_{a}^{(p),*}$;
		\ENDFOR
	\ENDFOR
\UNTIL{convergence}
\end{algorithmic}
\end{algorithm}

Finally, we also consider the case in which each AP uses a predetermined percentage of the available power to serve the UAVs. The rationale behind this approach is based on the fact that networks operators might want to guarantee a specific quality of service for UAVs and/or GUEs. We denote by $\kappa$ the fraction of the available power that each AP uses for the UAVs, and with ${\cal G}_a$ and ${\cal U}_a$ the sets of the GUEs and the UAVs served from the $a$-th AP. 

Problem \eqref{Prob:MinRateApp} is solved by properly choosing the blocks $\overline{\bbeta}_{a}^{(p)}, \; \forall p=1,\ldots, P_a, \,a=1,\ldots,N_A$ that contain only GUEs or only UAVs, using the constraints on the maximum power as $\left(1-\kappa\right) \eta^{\rm DL}_a$ and $\kappa \eta^{\rm DL}_a$, respectively and 
\begin{equation}
\rho_{a,k}^{\rm DL}=  \left\lbrace
\begin{array}{llll}
\ds  \left(\ds \sum_{j\in{\cal G}_a} {\gamma_{j,a}}\right)^{-1}, \; & \text{if} \; k \in \mathcal{G}_a \, ,
\\
\ds \left(\ds \sum_{j\in{\cal U}_a} {\gamma_{j,a}}\right)^{-1}, \; & \text{if} \; k \in \mathcal{U}_a.
\end{array} \right. 
\end{equation}
The full details are omitted for the sake of brevity.

\subsubsection{Waterfilling Power Allocation}

Next, we propose a heuristic power allocation policy inspired by the well-known waterfilling strategy \cite{Cover-Thomas}.
We assume that the ``noise'' level for the communication between the $a$-th AP and the $k$-th user is written as 
$
L_{k,a}= \frac{\sigma^2_z}{\gamma_{k,a}}$, so that the waterfilling power allocation (WFPA) strategy is expressed as
\begin{equation}
P_{k,a}^{\rm DL}=  \left\lbrace
\begin{array}{llll}
\left( \nu_a - L_{k,a} \right)^+, \; & \text{if} \; k \in \mathcal{K}_a \, ,
\\
0 & \text{otherwise}.
\end{array} \right. 
\label{WFPA_powers}
\end{equation}
where $\nu_a$ is the water level, $(\cdot)^+=\max\{0, \cdot\}$, and the following constraint holds:
\begin{equation}
\sum_{k \in \mathcal{K}_a}\left( \nu_a - L_{k,a} \right)^+ = \eta^{\rm DL}_a \, .
\end{equation}
This heuristic power allocation rule forces a CF deployment to behave like an AP-centric 
system, since Eq. \eqref{WFPA_powers} implicitly makes a selection of the users to serve.

Considering the case in which each AP uses a fraction $\kappa$ of the available power to serve the UAVs, the WFPA rule becomes
\begin{equation}
P_{k,a}^{\rm DL}=  \left\lbrace
\begin{array}{llll}
\left( \nu_a^{(GUE)} - L_{k,a} \right)^+, \; & \text{if} \; k \in \mathcal{G}_a \, ,
\\
\left( \nu_a^{(UAV)} - L_{k,a} \right)^+, \; & \text{if} \; k \in \mathcal{U}_a \, ,
\\
0 & \text{otherwise}.
\end{array} \right. 
\end{equation}
where $\nu_a^{(GUE)}$ and $\nu_a^{(UAV)}$ are the water levels for the GUEs and the UAVs, respectively, and the following constraints are to be fulfilled.
\begin{equation}
\begin{array}{llll}
\ds \sum_{k \in \mathcal{G}_a}\left( \nu_a^{(GUE)} - L_{k,a} \right)^+ = \left(1-\kappa\right)\eta^{\rm DL}_a \, ,
\\
\ds \sum_{k \in \mathcal{U}_a}\left( \nu_a^{(UAV)} - L_{k,a} \right)^+ = \kappa \eta^{\rm DL}_a.
\end{array}
\end{equation}

\subsubsection{Proportional Power Allocation}

As a baseline power allocation strategy, we also consider proportional power allocation (PPA). Letting $P_{k,a}^{\rm DL}=\eta_{k,a}^{\rm DL} \gamma_{k,a}$ denote the power transmitted by the $a$-th AP to the $k$-th user, we have the policy:
\begin{equation}
P_{k,a}^{\rm DL}=  \left\lbrace
\begin{array}{llll}
\ds \eta^{\rm DL}_a \frac{ \gamma_{k,a}}{\ds \sum_{j\in{\cal K}_a} {\gamma_{j,a}}}, \; & \text{if} \; k \in \mathcal{K}_a,
\\
0 & \text{otherwise}.
\end{array} \right.
\end{equation}
The above strategy is such that the generic $a$-th AP shares its power $\eta_a^{\rm DL}$ in a way that is proportional to the estimated channel strengths. In this way, users with good channel coefficients will receive a larger share of the transmit power than users with bad channels. 

Additionally, we consider the case in which each AP uses a fraction $\kappa$ of the available power to serve the UAVs. This is because UAVs tend to absorb a large share of the system resources in the case of proportional power allocation, due to the fact that they generally enjoy stronger channels than GUEs, as later demonstrated in the numerical results. Accordingly, the division of the power resources provides a further degree of flexibility for networks operators and facilitates guaranteeing the GUEs' performance. Following this approach, the PPA rule can be written as

\begin{equation}
P_{k,a}^{\rm DL}=  \left\lbrace
\begin{array}{llll}
\ds \left(1- \kappa\right) \eta^{\rm DL}_a \frac{\gamma_{k,a}}{\ds \sum_{j\in{\cal G}_a} {\gamma_{j,a}}}, \; & \text{if} \; k \in \mathcal{G}_a \, ,
\\
\ds \kappa \eta^{\rm DL}_a \frac{ \gamma_{k,a}}{\ds \sum_{j\in{\cal U}_a} {\gamma_{j,a}}}, \; & \text{if} \; k \in \mathcal{U}_a \, ,
\\
0 & \text{otherwise} \, .
\end{array} \right. 
\end{equation}

\subsection{Uplink Data Transmission} \label{Power_allocation_section_U}

\subsubsection{Minimum-rate Maximization}
Also for the uplink, a power allocation strategy based on the minimum-rate maximization can be conceived. We are thus faced with the optimization problem

\begin{subequations}\label{ProbMinRate_UL}
\begin{align}
&\ds\max_{\widetilde{\bbeta}}\;\min_{k\in{\cal K}}\; \mathcal{R}_{k}^{\rm UL}(\widetilde{\bbeta})\label{Prob:aMinRate_UL}\\
&\;\textrm{s.t.}\;0 \leq \eta_{k,a}^{\rm UL}\leq P_{{\rm max},k}\;,\forall\;k=1,\ldots,K,\label{Prob:bMinRate_UL}
\end{align}
\end{subequations}
where $\mathcal{R}_{k}^{\rm UL}(\widetilde{\bbeta})= W \text{SE}_{k, {\rm UB}}^{\rm UL}$ is the achievable uplink rate and $\widetilde{\bbeta}$ is the $K\times 1$ vector collecting the uplink transmit powers of all the users. Using similar arguments as done for the downlink and defining the $R$-dimensional variable blocks $\widetilde{\bbeta}^{(q)}$, $q=1,\ldots,Q$, collecting the $q$-th block of uplink transmit powers, the minimum-rate maximization with respect to the variable block $\widetilde{\bbeta}^{(q)}$ is cast as
\begin{subequations}\label{Prob:MinRate_ULSub}
\begin{align}
&\ds\max_{\widetilde{\bbeta}^{(q)},t}\;t\label{Prob:aMinRate_ULSub}\\
&\;\textrm{s.t.}\;0 \leq \eta_{k}^{\rm UL}\leq P_{{\rm max},k}\;,\forall\;k=1,\ldots,K.\label{Prob:bMinRate_ULSub}\\
&\;\;\;\quad  \mathcal{R}_{k}^{\rm UL}(\widetilde{\bbeta}^{(q)},\widetilde{\bbeta}^{(-q)})\geq t, \; \forall \; k=1,\ldots, K.\label{Prob:cMinRate_ULSub}
\end{align}
\end{subequations}
Letting $\widetilde{\mathcal{R}}_k^{\rm UL}$ be a suitable upper bound to  $\mathcal{R}_k^{\rm UL}$, following a similar approach as in Section \ref{Max_Min_PA_DL}, Problem \eqref{Prob:MinRate_ULSub} can be tackled by the sequential optimization framework, by defining the $q$-th problem of the sequence:
\begin{subequations}\label{Prob:MinRate_ULApp}
\begin{align}
\mathcal{P}_q :&\ds\max_{\widetilde{\bbeta}^{(q)},t}\;t\label{Prob:aMinRate_ULApp}\\
&\;\textrm{s.t.}\;0 \leq \eta_{k}^{\rm UL}\leq P_{{\rm max},k}\;,\forall\;k=1,\ldots,K.\label{Prob:bMinRate_ULApp}\\
&\;\;\;\quad  \widetilde{\mathcal{R}}_{k}^{\rm UL}\left(\widetilde{\bbeta}^{(q)},\widetilde{\bbeta}_{0}^{(q)},\widetilde{\bbeta}^{(-q)}\right)\geq t, \; \forall \; k=1,\ldots, K.\label{Prob:cMinRate_ULApp}
\end{align}
\end{subequations}
The resulting power control procedure can be stated as in Algorithm \ref{Alg:MRMax_UL}. 
\begin{algorithm}[t]
\caption{Minimum-rate maximization in uplink}
\begin{algorithmic}[1]
\label{Alg:MRMax_UL}
\STATE Set $i=0$ and choose any feasible $\widetilde{\bbeta}^{(2)},\ldots, \widetilde{\bbeta}^{(Q)}$;
\REPEAT

	\FOR{$q=1\to Q$} 
	\REPEAT
	\STATE Choose any feasible $\widetilde{\bbeta}_{0}^{(q)}$;
	\STATE Let $\widetilde{\bbeta}^{(q),*}$ be the solution of $\mathcal{P}_q$ in\eqref{Prob:MinRate_ULApp};
	\STATE $\widetilde{\bbeta}_{0}^{(q)}=\widetilde{\bbeta}^{(q),*}$;
	\UNTIL{convergence}
	\STATE $\widetilde{\bbeta}^{(q)}=\widetilde{\bbeta}^{(q),*}$;
	\ENDFOR
\UNTIL{convergence}
\end{algorithmic}
\end{algorithm}

\subsubsection{Fractional Power Control}

We adopt fractional power control (FPC) as the reference power adjustment rule for uplink data transmission \cite{BarGalGar2018GC, 3GPP_36777}. With FPC, the transmit power of the $k$-th user can be expressed as $
\eta_k^{\rm UL}= \text{min} \left( P_{{\rm max},k}, P_0 \zeta_k^{-\alpha}\right)$, 
where $P_{{\rm max},k}$ is the maximum $k$-th user transmit power, and $P_0$ is a cell-specific parameter configurable by the serving AP, $\alpha$ is a path loss compensation factor. Moreover, $\zeta_k$ captures the large scale fading that the $k$-th user experiences to the serving APs in $\mathcal{A}_k$, and is obtained as
$
\zeta_k= \sqrt{\sum_{a \in \mathcal{A}_k} {\text{tr}\left( \mathbf{G}_{k,a} \right)}}.
$

\section{NUMERICAL RESULTS AND KEY INSIGHTS} \label{Numerical_results}
\begin{table}
\centering
\caption{Cell-free system parameters}
\label{table:parameters}
\def\arraystretch{1.2}
\begin{tabulary}{\columnwidth}{ |p{2.5cm}|p{4.8cm}| }
\hline
	\textbf{Deployment} 			&  \\ \hline
  	AP distribution				&  Horizontal: uniform, vertical: 10~m \\ \hline
  GUE distribution 				& Horizontal: uniform, vertical: 1.65~m\\ \hline
	UAV distribution 				& Horizontal: uniform, vertical uniform between 22.5~m and 300~m \cite{3GPP_36777}\\ \hline \hline
	\textbf{PHY and MAC} 			&  \\ \hline
	Carrier frequency, bandwidth		&  $f_0=1.9$ GHz, $W = 20$ MHz \\ \hline
	AP antenna array			& Four-element ULA with $\lambda/2$ spacing\\ \hline
	User antennas 		& Omnidirectional with 0~dBi gain\\ \hline
	\multirow{2}{*}{Power allocation}		& DL: proportional power allocation (PPA), waterfilling power allocation (WFPA), or minimum-rate maximization power allocation (MR max) \\ \cline{2-2}
	& UL: FPC with $\alpha=0.5$ and $P_0=-10$ dBm, or minimum-rate maximization power allocation (MR max)  \\ \hline
	Thermal noise 				& -174 dBm/Hz spectral density \\ \hline
	Noise figure 			& 9 dB at APs/GUEs/UAVs \\ \hline
	User association		& Cell-free (CF) or user centric (UC) \\ \hline
	Traffic model		& Full buffer \\ \hline
\end{tabulary}
\end{table}

The simulation setup for the numerical results is detailed in the following. We consider a square area of 1 km$^2$ wrapped around at the edges to avoid boundary effects. In this scenario, we evaluate the data rates per user, obtained as the product of the spectral efficiency by the system bandwidth $W$, of two different network topologies according to the number and characteristics of the APs deployed:
\begin{enumerate}
\item CF and UC architectures with $N_\mathrm{A}=100$ APs comprised of $N_{\rm AP} = 4$ antennas each. The maximum downlink power transmitted by the $a$-th AP is $\eta^{\rm DL}_a= 200$~mW, $\forall a \in \mathcal{A}$.
\item As a benchmarking network structure, a multi-cell massive MIMO (mMIMO) system with $N_\mathrm{BS}=4$ BSs with $N_{\rm BS} = 100$ antennas each. So that the overall downlink transmit power is kept constant w.r.t.\ the CF and UC architectures, we consider that the maximum downlink power per mMIMO BS is $\eta^{\rm DL}_a= 5$~W. In order to consider a fair comparison, we assume also in the case of mMIMO system a matched filtering for both the uplink and downlink.

\end{enumerate}

With regard to the channels from GUEs to the APs, i.e., when $k \in \mathcal{G}$, we consider a urban environment with a high density of buildings and obstacles where all the GUEs are in NLOS, i.e., $p_{\rm LOS}\left(d_{{\rm 2D},k,a}\right)=0, \; \forall k \in \mathcal{G}$. The large scale coefficient $\beta_{k,a}$ in dB is modelled as in \cite[Table B.1.2.2.1-1]{3GPP_36814_GUE_model}, i.e.:
$
\beta_{k,a} [\text{dB}]=-36.7\log_{10}(d_{k,a})-22.7-26\log_{10}(f)+z_{k,a},
$
where $z_{k,a} \sim \mathcal{N}\left( 0, \sigma_{\rm sh}^2\right)$ represents the shadow fading. The shadow fading coefficients from an AP to different GUEs are correlated as in \cite[Table B.1.2.2.1-4]{3GPP_36814_GUE_model}. Instead, the shadow fading correlation among GUEs follows \cite{bjornson2019CF_MMSE} 
\begin{equation}
\mathbb{E}[z_{k,a}z_{j,b}]= \left\lbrace
\begin{array}{llll}
&\sigma_{\rm sh}^2 2^{-\frac{\rho_{k,j}}{d_0}} \;, & a=b, \\
&0 \; ,& a \neq b \, ,
\end{array} \right.
\end{equation}
where $\rho_{k,j}$ is the distance between the $k$-th and the $j$-th GUEs, $d_0=9$ m, and $\sigma_{\rm sh}=4$.

When considering the channels between the UAVs and the APs, i.e., when $k \in \mathcal{U}$, we evaluate the LOS probability as specified in \cite[Table B-1]{3GPP_36777}. Similarly, the large scale fading $\beta_{k,a} [\text{dB}]$ is evaluated following \cite[Table B-2]{3GPP_36777}.

To understand the impact that UAVs have on these cellular networks, we compare a scenario with $N_{\rm G}=60$ GUEs and no UAVs, with a scenario with $N_{\rm G}=48$ GUEs and $N_{\rm U}=12$ UAVs. In this setup, we consider $\tau_c = 200$ time/frequency samples, corresponding to a coherence bandwidth of $200$ kHz and a coherence time of $1$~ms \cite{Ngo_CellFree2016}. Equal uplink/downlink split of the available time/frequency resources after training is assumed, i.e., $\tau_{\rm d}=\tau_{\rm u}=\frac{\tau_c-\tau_p}{2}$. We assume that the length of the uplink pilot training sequences is $\tau_p=32$, we take a set $\mathcal{P}_{\tau_p}$ of orthogonal pilots with length $\tau_p$ and randomly assign the pilot sequences in $\mathcal{P}_{\tau_p}$ to the GUEs and UAVs in the system, i.e., our results account for the impact of pilot contamination. The uplink transmit power during training is $\eta_k=\tau_p \overline{\eta}_k$, with $\overline{\eta}_k=100$~mW $ \forall k \in \mathcal{K}$. During uplink data transmission, the maximum uplink power transmitted by the $k$-th user is $P_{\rm max}^{\rm UL}=100$ mW, $\forall k \in \mathcal{K}$. The remaining system parameters are detailed in Table \ref{table:parameters}.


\begin{figure}[!t]
\centering
\includegraphics[scale=0.48]{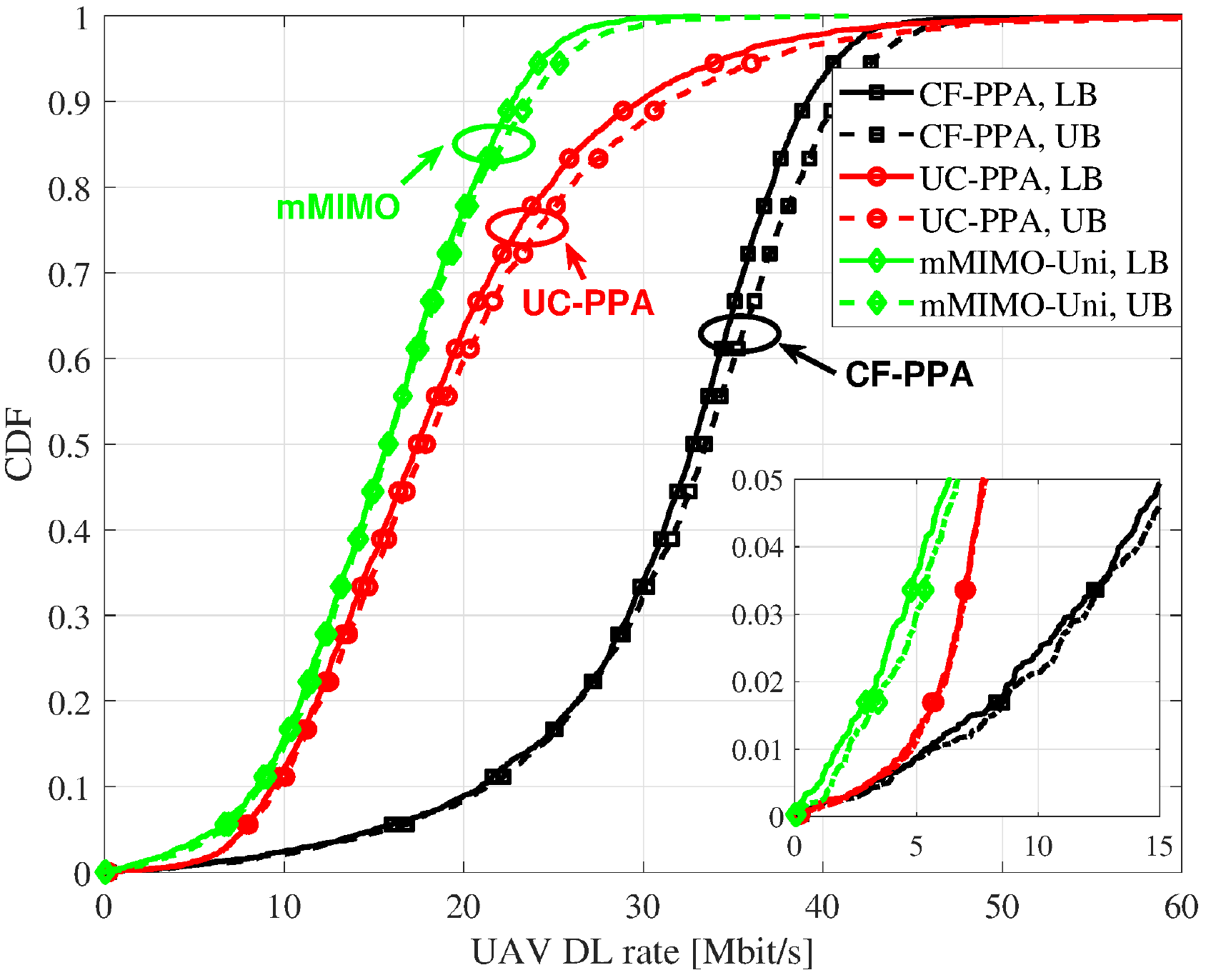}
\caption{DL rates for UAVs under: (i) cell-free with proportional power allocation (CF-PPA), (ii) user-centric with $A_k=10$ and proportional power allocation (UC-PPA), and (iii) multi-cell mMIMO with uniform power (mMIMO-Uni).}
\label{Fig:UAV_DL}
\end{figure}

\subsection{Downlink performance}

We start by turning our attention to the downlink performance.

\begin{figure*}[!t]
\centering
\includegraphics[scale=0.5]{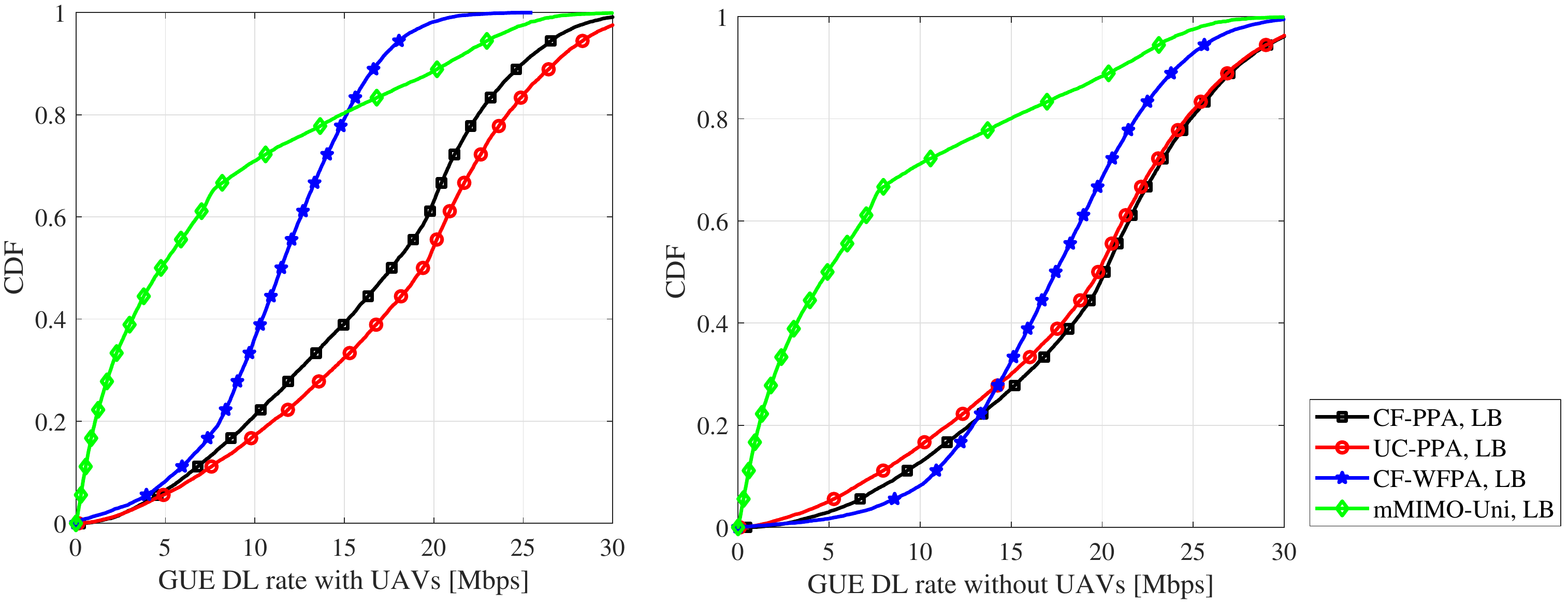}
\caption{DL rates for GUEs in scenarios with/without UAVs under: (i) cell-free with proportional power allocation (CF-PPA), (ii) user-centric with $A_k=10$ and proportional power allocation (UC-PPA), (iii) cell-free with waterfilling power allocation (CF-WFPA), and (iv) multi-cell mMIMO with uniform power (mMIMO-Uni).}
\label{Fig:GUE_DL}
\end{figure*}

\begin{figure}[!t]
\centering
\includegraphics[scale=0.5]{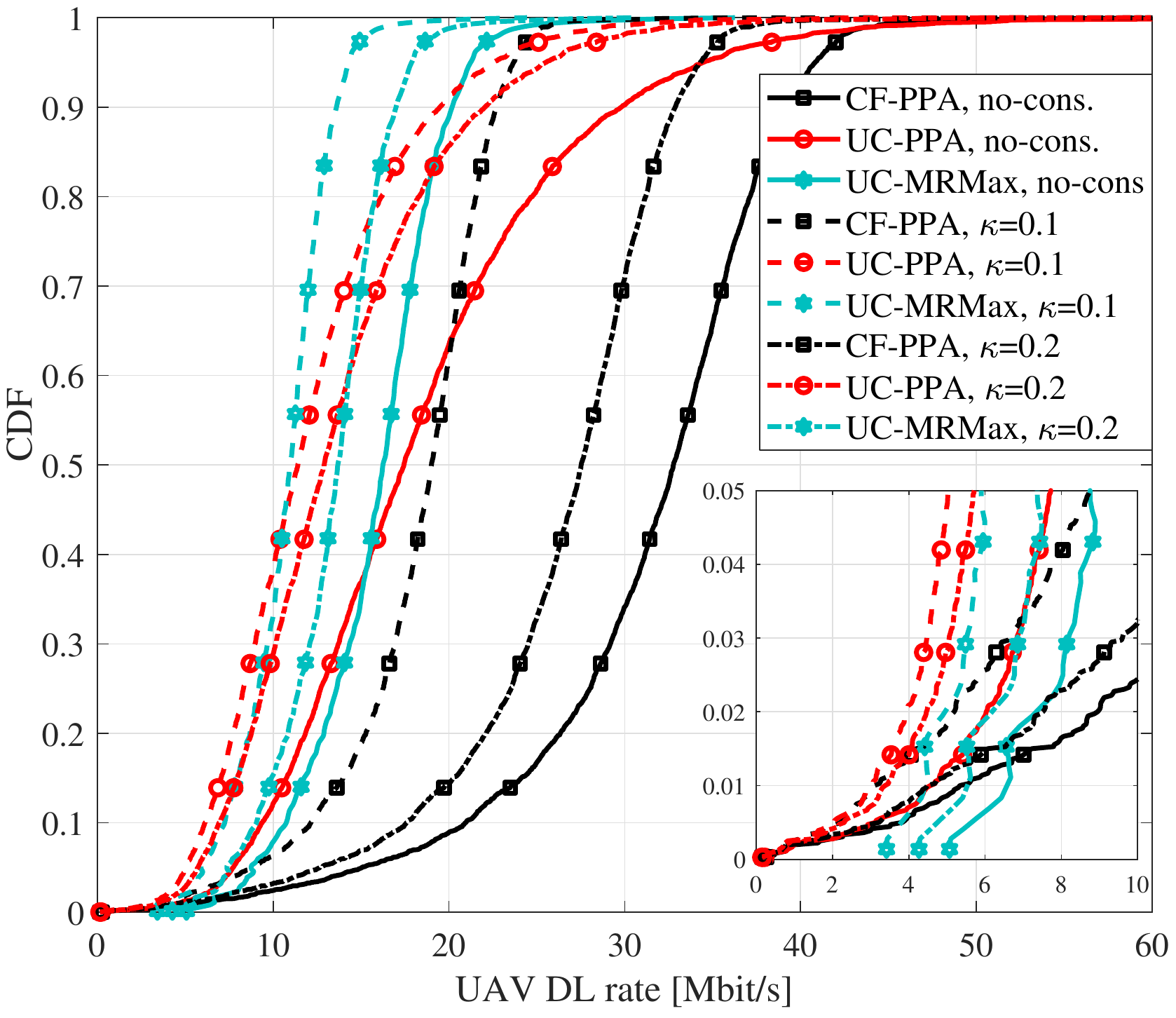}
\caption{UAV DL rates for deployments with unconstrained power allocation per UAV (no-cons.) and deployments dedicating a fixed power share $\kappa$ to UAVs. The minimum rate maximizing power allocation is reported for the user-centric scheme with $A_k=10$ (UC-MR max).}
\label{Fig:MRmax_DL_UAVs}
\end{figure}

\begin{figure}[!t]
\centering
\includegraphics[scale=0.5]{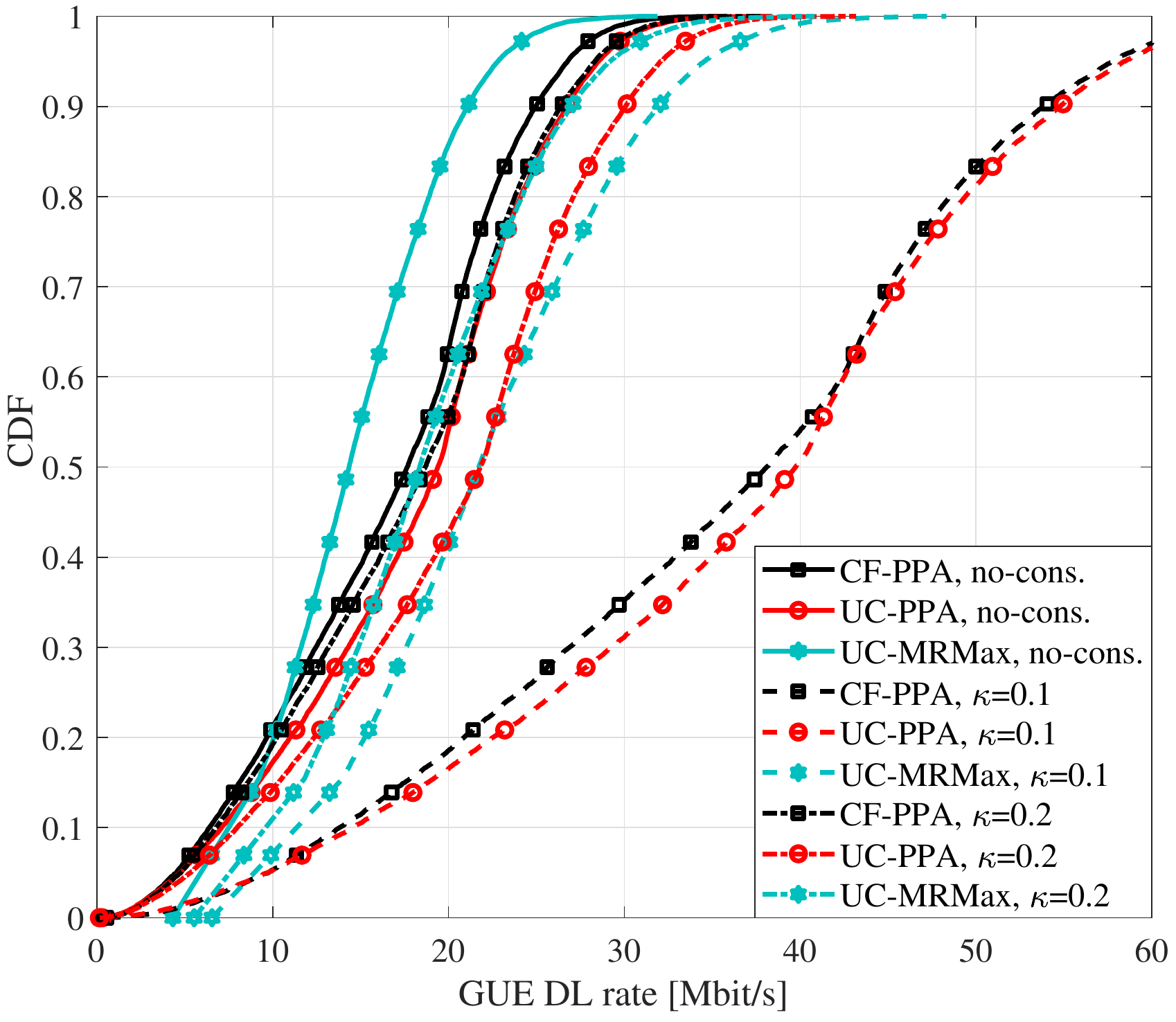}
\caption{GUE DL rates for deployments with unconstrained power allocation per UAV (no-cons.) and deployments dedicating a fixed power share $\kappa$ to UAVs. The minimum rate maximizing power allocation is reported for the user-centric scheme with $A_k=10$ (UC-MR max).}
\label{Fig:MRmax_DL_GUEs}
\end{figure}

Fig. \ref{Fig:UAV_DL} reports the cumulative distribution functions (CDFs) of the DL UAV rates for the following network deployments: (i) a CF architecture with PPA (CF-PPA), (ii) a UC deployment with $A_k=10$ and PPA (UC-PPA), and (iii) multi-cell mMIMO deployment with uniform power allocation (mMIMO-Uni). The results of Fig. \ref{Fig:UAV_DL} demonstrate the tightness of the lower and upper downlink spectral efficiency bounds derived in Sec.\ \ref{DL_Bounds} for the case of UAVs. Moreover, Fig. \ref{Fig:UAV_DL} illustrates that both CF and UC architectures outperform the considered multi-cell mMIMO deployment, which is consistent with the consideration that inter-cell interference can greatly harm the performance of multi-cell mMIMO systems \cite{GarGerLop2018,GeraciUAVs_Access2018}. Remarkably, Fig. \ref{Fig:UAV_DL} also shows that the CF-PPA architecture can provide substantially larger UAV rates than the UC-PPA deployment. This is because UAVs experience good propagation conditions with a large number of ground BSs simultaneously, and therefore the reduced number of serving BSs of the UC-PPA deployment leads to a substantial performance degradation when compared to the CF-PPA architecture---where all BSs can communicate with the UAVs.

Fig. \ref{Fig:GUE_DL} represents the lower bounds of the DL GUE rates for the same network deployments of Fig. \ref{Fig:UAV_DL} and, additionally, a CF architecture with WFPA (CF-WFPA). The results of Fig. \ref{Fig:GUE_DL} illustrate that, while the mMIMO deployment generally provides for GUEs a worse performance than the competing schemes, it approximately preserves such performance under the presence of UAVs---thanks to the effective mitigation of the highly directional UAV-generated uplink pilot contamination through LMMSE channel estimation \cite{GarGerLop2018}. Differently to the UAV behaviour described in Fig. \ref{Fig:UAV_DL}, we can also observe that there are no substantial performance differences between the CF and UC schemes, since GUEs experience good propagation conditions with a limited number of ground BSs simultaneously. Interestingly, Fig. \ref{Fig:GUE_DL} shows that the GUE rates are substantially degraded when UAVs are present in the network for the case of CF-WFPA (from approximately 17 Mbit/s to about 10 Mbit/s in median). This is because the WFPA allocates more power to the users with largest channel coefficients, which leads UAVs to take a considerable share of the system resources and is the main reason behind the introduction of the power control rules that guarantee a fixed power share to GUEs in Sec.\ \ref{Power_allocation_section_D}.

Fig. \ref{Fig:MRmax_DL_UAVs} and \ref{Fig:MRmax_DL_GUEs} are devoted to assess the impact that both 1) the power control rule that maximizes the minimum rate, and 2) of the strategy that constrains the power share reserved to the UAVs have on the UAV and GUE rates, respectively. Specifically, these figures represent the CDF of the LB rate per user for the CF and UC deployments dedicating a fixed power share $\kappa=0.1$ and $\kappa=0.2$ to UAVs, and with unconstrained power allocation per UAV (no-cons.).

The results of Fig. \ref{Fig:MRmax_DL_UAVs} corroborate both 1) the effectiveness of the power control rule that maximizes the minimum rate---as shown in the lower part of the zoomed area---, and 2) that limiting the value of $\kappa$ negatively impacts the performance of the UAVs. Instead, Fig. \ref{Fig:MRmax_DL_UAVs} demonstrates that the GUE rates can greatly benefit from constraining the power dedicated to UAVs. Overall, the trends of Fig. \ref{Fig:MRmax_DL_UAVs} and \ref{Fig:MRmax_DL_GUEs} illustrate the importance of properly optimizing $\kappa$ to provide an adequate performance to both UAVs and GUEs. 
 
  
\subsection{Uplink performance}

\begin{figure}[!t]
\centering
\includegraphics[scale=0.5]{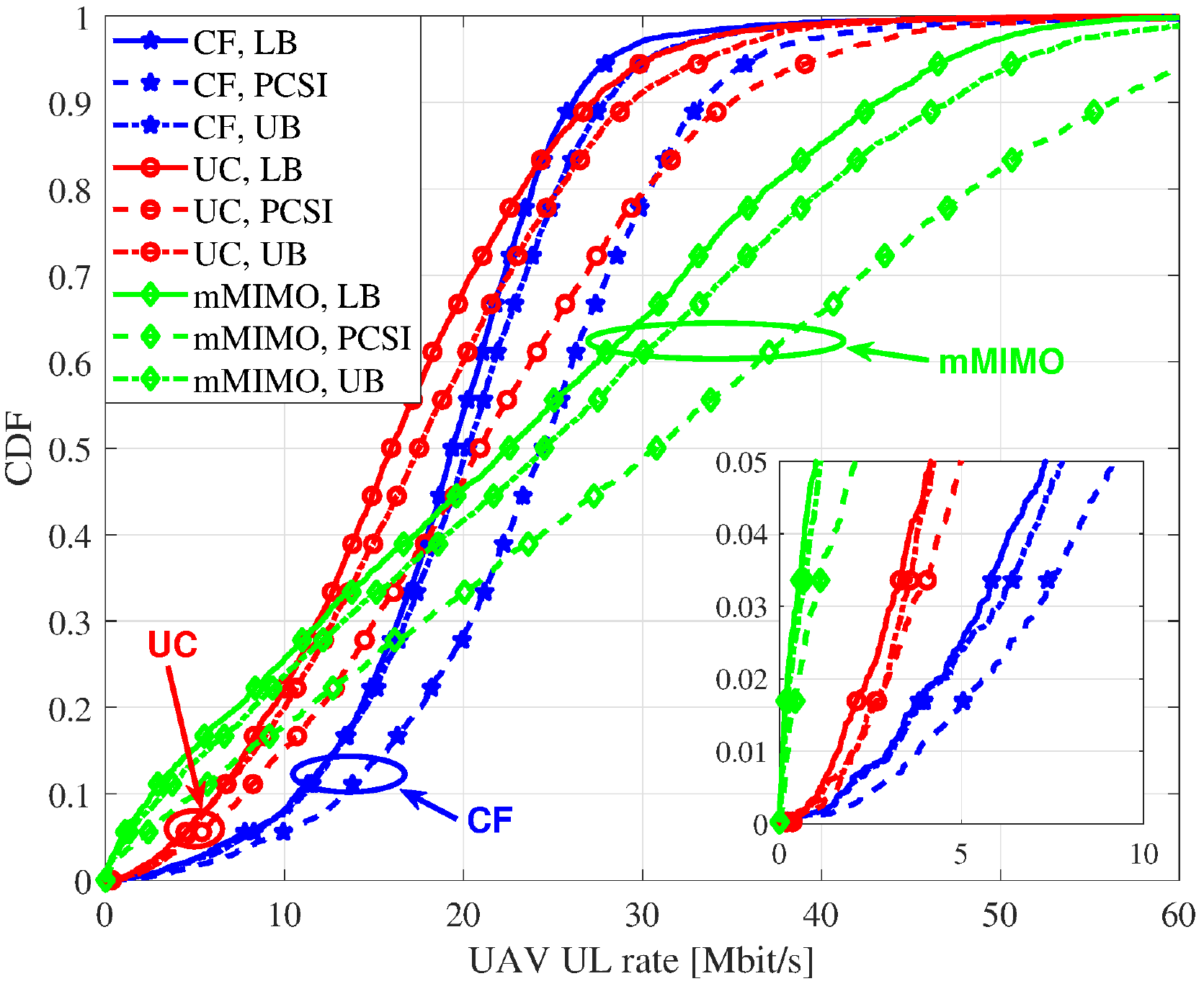}
\caption{UL rates for UAVs under: (i) cell-free (CF), (ii) user-centric (UC) with $A_k=10$, and (iii) multi-cell mMIMO (mMIMO) approaches.}
\label{Fig:UAV_UL}
\end{figure}

We continue by summarizing the uplink performance results. Fig. \ref{Fig:UAV_UL} represents the CDFs of the UL UAV rates in (i) a CF architecture with FPC; (ii) a UC architecture with FPC, and (iii) a benchmark multi-cell mMIMO deployment with FPC. Inspecting this figure, it can be concluded that there exists a UAV performance trade-off between CF/UC architectures and mMIMO deployments:
\begin{itemize}
\item The CF/UC architectures provide substantial performance gains over the baseline mMIMO deployment for the worst-performing UAVs located in the lower part of the CDFs. For instance, the LB of the UAV rates at the 5-th percentile grows from approximately 1 Mbit/s with mMIMO to 7.3 Mbit/s with the CF deployment. This improvement can be explained by noticing that UAVs served by CF/UC architectures do not experience the cell-edge problems that occur with conventional mMIMO deployments. Similarly to the trends of Fig. \ref{Fig:UAV_DL}, the CF architecture generally provides better UAV rates than the UC deployments, since UAVs benefits from transmitting their data to the large number of APs with good propagation conditions.
\item Instead, the benchmark multi-cell mMIMO deployment clearly outperforms CF/UC architectures for the best-performing UAVs located in the upper part of the CDFs. This is because mMIMO BSs can provide substantial signal power gains for those UAVs located close to them. 
\end{itemize} 
The results of Fig. \ref{Fig:UAV_UL} also illustrate that the performance impact of having imperfect CSI is similar for all considered deployments.

\begin{figure*}[!t]
\centering
\includegraphics[scale=0.5]{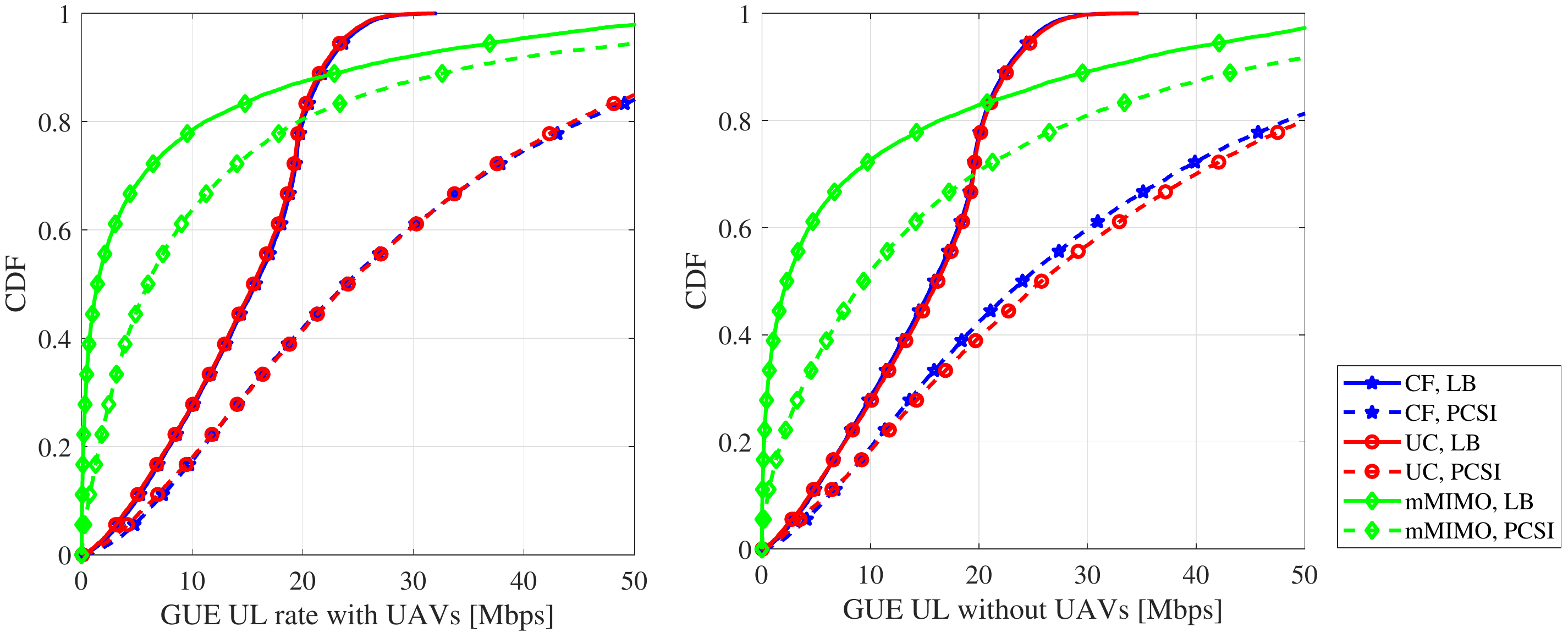}
\caption{UL rates for GUEs in scenarios with/without UAVs under: (i) cell-free (CF), (ii) user-centric (UC) with $A_k=10$, and (iii) multi-cell mMIMO (mMIMO) approaches.}
\label{Fig:GUE_UL}
\end{figure*}

Fig. \ref{Fig:GUE_UL} includes the CDFs of the UL GUE rates for scenarios with and without UAVs, and the same deployments considered in Fig. \ref{Fig:UAV_UL}. Interestingly, the results of Fig. \ref{Fig:GUE_UL} illustrate that---differently to what occurs in the baseline mMIMO deployment---the introduction of UAVs in the network has a negligible impact on the performance of both CF and UC architectures. This is because a) the UAV uplink pilot contamination is adequately managed by the LMMSE channel estimator, and b) CF and UC architectures spread the serving APs in wider areas, which facilitates the spatial separation of the incoming---highly directional---UAV signals from those transmitted by GUEs. Consistently with the results obtained in \cite{buzziWCL2017, Buzzi_WSA2017}, Fig. \ref{Fig:GUE_UL} also corroborates that GUEs can achieve similar performances in CF and UC deployments.

\begin{figure}[!t]
\centering
\includegraphics[scale=0.55]{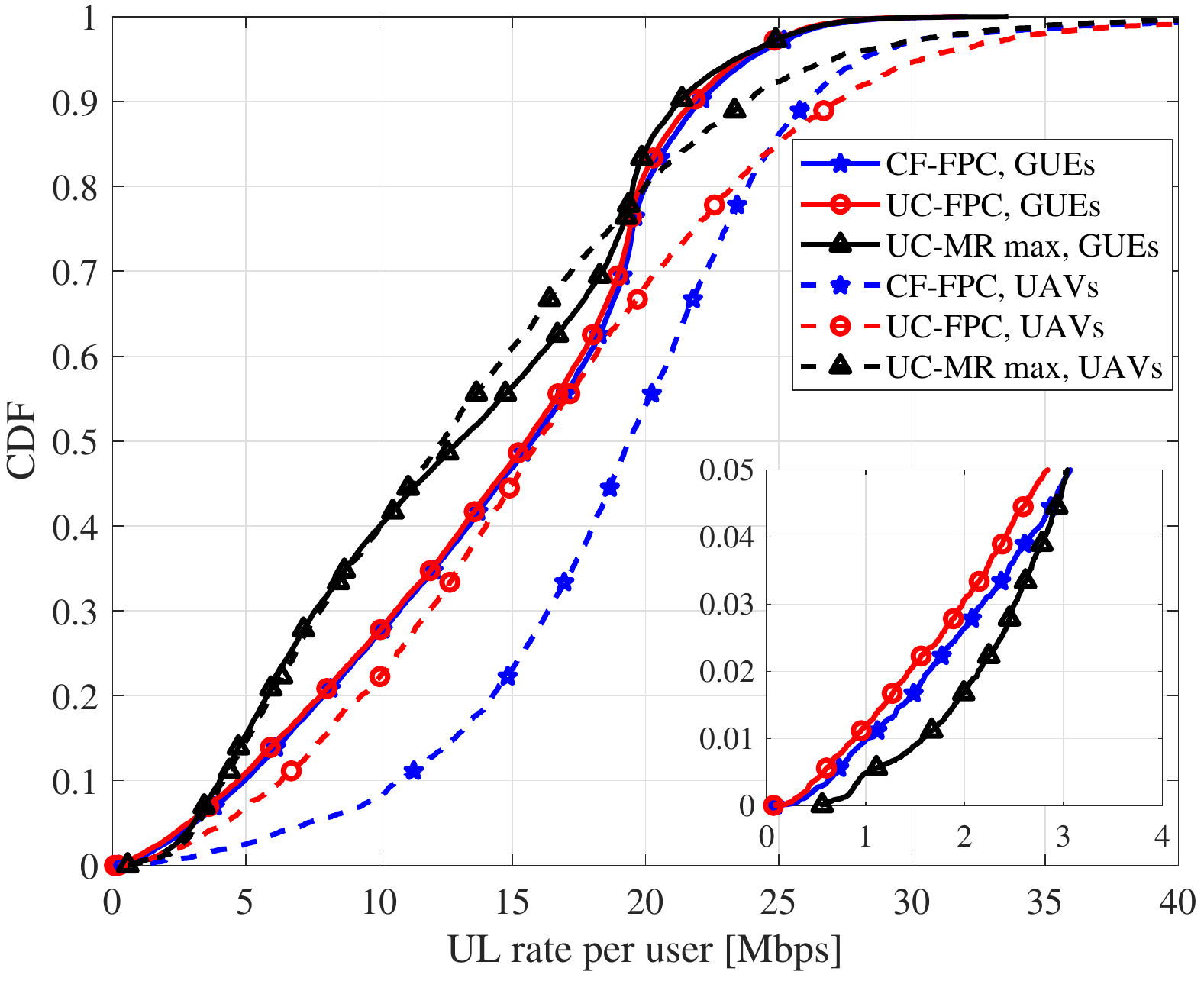}
\caption{UL rates for GUEs and UAVs with FPC and minimum-rate maximization power allocation (MR max) under: (i) cell-free (CF), and (ii) user-centric (UC) with $A_k=10$.}
\label{Fig:MRmax_UL}
\end{figure} 
 
Finally, Fig. \ref{Fig:MRmax_UL}, is devoted to the performance assessment of the power control rule maximizing the minimum UL rates. This figure reports the CDFs of the GUE and UAV UL LB rates in the UC and CF  scenarios. Comparing the results of UC-FPC with UC-MR max, we can observe that the 99$\%$-likely per GUE rates UC deployment increase from approximately 0.94 Mbit/s (UC-FPC) to 1.5 Mbit/s (+ 60$\%$), thus confirming the effectiveness of the proposed strategy for improving the system fairness across users, and in reducing the performance unbalance between GUEs and UAVs.

\section{CONCLUSIONS} \label{conclusions_section}
This paper has investigated  the use of CF and UC architectures for supporting wireless communications with UAVs. Assuming that the propagation channel between the users, either UAVs or GUEs, and the APs follows a Ricean distribution, closed form formulas for the achievable spectral efficiency LB for uplink and downlink with LMMSE channel estimation have been derived. 
Several power control rules have been considered, including one maximizing the minimum-rate maximizing power allocation strategy, based on sequential lower bound maximization. Numerical results have show that CF and its low-complexity UC alternative may provide superior performance in the support of UAVs communications than traditional multi-cell mMIMO deployments. Moreover, since UAVs generally enjoy better channel conditions than GUEs, it may be desirable to constrain the share of power that the APs should use to serve the former.

\section*{Appendix}

\subsection{Derivation of Lemma 1}
\label{Appendix_DL}

\noindent
To evaluate the closed-form expression for the downlink spectral efficiency in Eq. \eqref{eq:SE_general}, we need to compute $D_{k}$, $\mathbb{E}\left[ |B_{k}|^2\right]$ and $\mathbb{E}\left[ |I_{k,j}|^2\right]$.

\begin{enumerate}
\item \textit{Computation of} $D_{k}$: Denoting by $\widetilde{\mathbf{g}}_{k,a}= \mathbf{g}_{k,a} -\widehat{\mathbf{g}}_{k,a}$ the channel estimation error, the well-known LMMSE estimation property results in the fact that $\widetilde{\mathbf{g}}_{k,a}$ and $\widehat{\mathbf{g}}_{k,a}$ are independent. Using the independence between $\widetilde{\mathbf{g}}_{k,a}$ and $\widehat{\mathbf{g}}_{k,a}$, and substituting $\mathbf{g}_{k,a}=\widehat{\mathbf{g}}_{k,a}+\widetilde{\mathbf{g}}_{k,a}$,  we have

\begin{equation}
\begin{array}{lll}
{D}_{k}&=\mathbb{E}\left[ \ds \sum_{a\in{\cal A}_k} \ds \sqrt{\eta_{k,a}^{\rm DL}} \left( \widehat{\mathbf{g}}_{k,a}+ \widetilde{\mathbf{g}}_{k,a}\right)^H  \widehat{\mathbf{g}}_{k,a}\right]  \\  & = \ds \sum_{a\in{\cal A}_k} \ds \sqrt{\eta_{k,a}^{\rm DL}} \mathbb{E}\left[  \widehat{\mathbf{g}}_{k,a}^H \widehat{\mathbf{g}}_{k,a} \right]=  \ds \sum_{a\in{\cal A}_k} \ds \sqrt{\eta_{k,a}^{\rm DL}} \gamma_{k,a} \; .
\end{array}
\label{Dk}
\end{equation}

\item \textit{Computation of} $\mathbb{E}\left[ |B_{k}|^2\right]$: Since the variance of a sum of independent RVs is equal to the sum of their variances, we have
\begin{equation}
\begin{array}{llll}
&\mathbb{E}\left[ |{B}_{k}|^2\right] = \ds \sum_{a\in{\cal A}_k} \ds \eta_{k,a}^{\rm DL} \mathbb{E}\left[ \left| \mathbf{g}_{k,a}^H  \widehat{\mathbf{g}}_{k,a} - \mathbb{E}\left[ \mathbf{g}_{k,a}^H  \widehat{\mathbf{g}}_{k,a} \right]\right|^2 \right] \\ & = \ds \sum_{a\in{\cal A}_k} \eta_{k,a}^{\rm DL} \left( \mathbb{E}\left[ \left| \mathbf{g}_{k,a}^H  \widehat{\mathbf{g}}_{k,a} \right|^2 \right] - \left|\mathbb{E}\left[ \mathbf{g}_{k,a}^H  \widehat{\mathbf{g}}_{k,a} \right]\right|^2 \right).
\end{array}
\label{E_Bk}
\end{equation}
We evaluate the mean $\mathbb{E}\left[ \left| \mathbf{g}_{k,a}^H  \widehat{\mathbf{g}}_{k,a} \right|^2 \right]$ using the definitions in Section \ref{MMSE_Ch_est}, which leads to

\begin{equation}
\begin{array}{llll}
\mathbb{E}\left[ \left| \mathbf{g}_{k,a}^H  \widehat{\mathbf{g}}_{k,a} \right|^2 \right]&=\mathbb{E}\left[ \left| \mathbf{g}_{k,a}^H  \mathbf{D}_{k,a}  \right. \right. \\ & \left. \left. \times \left( \ds \sum_{i \in \mathcal{K}} { \sqrt{ \eta_i} \mathbf{g}_{i,a} \boldsymbol{\phi}_i^H\boldsymbol{\phi}_k }+ \widetilde{\mathbf{w}}_{k,a} \right) \right|^2 \right]\, ,
\end{array}
\end{equation}
where we have $\widetilde{\mathbf{w}}_{k,a}=\mathbf{W}_a \boldsymbol{\phi}_k$ with i.i.d $ \mathcal{CN}\left(0,\sigma^2_w\right)$ entries. Using the assumptions that a) $\widetilde{\mathbf{w}}_{k,a}$ is independent from $\mathbf{g}_{i,a} \; \forall i$, the channels from different users are independent, and b) the following relation for complex Gaussian vectors
\begin{equation}
\mathbb{E}\left[ \left| \mathbf{g}_{k,a}^H  \mathbf{D}_{k,a} \mathbf{g}_{k,a} \right|^2\right]=\delta_{k,a}^{(k)} + \text{tr} \left(\mathbf{D}_{k,a}\mathbf{G}_{k,a}\mathbf{D}_{k,a}^H \mathbf{G}_{k,a}\right) \, ,
\label{relation_fourth_moment}
\end{equation}
where $\delta_{k,a}^{(k)}$ is defined as in Eq. \eqref{delta_ka_j}, we obtain with ordinary efforts

\begin{equation}
\begin{array}{llll}
\mathbb{E}\left[ \left| \mathbf{g}_{k,a}^H  \widehat{\mathbf{g}}_{k,a} \right|^2 \right]= \eta_k \delta_{k,a}^{(k)}+ \sqrt{\eta_k} \text{tr} \left(\mathbf{G}_{k,a}\mathbf{D}_{k,a}^H \mathbf{G}_{k,a}\right).
\end{array} 
\label{mean_Bk}
\end{equation}

Finally, substituting \eqref{mean_Bk} into Eq. \eqref{E_Bk}, we obtain
\begin{equation}
\begin{array}{lll}
\mathbb{E}\left[ |{B}_{k}|^2\right] =& \ds \sum_{a\in{\cal A}_k} \ds \eta_{k,a}^{\rm DL} \left( \eta_k \delta_{k,a}^{(k)} \right. \\ & \left. + \sqrt{\eta_k} \text{tr} \left(\mathbf{G}_{k,a}\mathbf{D}_{k,a}^H \mathbf{G}_{k,a}\right)-\gamma_{k,a}^2\right).
\end{array}
\label{Bk_final}
\end{equation}

\item \textit{Computation of} $\ds \mathbb{E}\left[ |{I}_{k,j}|^2\right]$: Using a similar approach as in Eq. \eqref{E_Bk} we obtain

\begin{equation}
\begin{array}{llll}
& \mathbb{E}\left[ |{I}_{k,j}|^2\right] =\eta_k \mathbb{E}\left[ \left| \ds \sum_{a\in{\cal A}_j} {\sqrt{\eta_{j,a}^{\rm DL}} \mathbf{g}_{k,a}^H \mathbf{D}_{j,a}\mathbf{g}_{k,a} }\right|^2  \right] \left|\boldsymbol{\phi}_k^H\boldsymbol{\phi}_j\right|^2 \\ & +  \ds \sum_{a\in{\cal A}_j} \ds\sum_{i \in \mathcal{K} \backslash k} \eta_{j,a}^{\rm DL} \eta_i  \text{tr} \left( \mathbf{D}_{j,a}\mathbf{G}_{i,a}\mathbf{D}_{j,a}^H \mathbf{G}_{k,a}\right) \left|\boldsymbol{\phi}_i^H\boldsymbol{\phi}_j\right|^2  \\ & + \sigma^2_w  \ds \sum_{a\in{\cal A}_j} \eta_{j,a}^{\rm DL} \text{tr} \left( \mathbf{D}_{j,a}\mathbf{D}_{j,a}^H \mathbf{G}_{k,a}\right).
\end{array}
\label{Interference_all2}
\end{equation}

The expectation that appears in Eq. \eqref{Interference_all2} can be computed, using a similar approach as in Eq. \eqref{mean_Bk}, as
\begin{equation}
\begin{array}{llll}
&\mathbb{E}\left[ \left| \ds \sum_{a\in{\cal A}_j} {\sqrt{\eta_{j,a}^{\rm DL}} \mathbf{g}_{k,a}^H \mathbf{D}_{j,a}\mathbf{g}_{k,a} }\right|^2  \right]   \\ & = \ds \sum_{a\in{\cal A}_j} {\eta_{j,a}^{\rm DL} \left( \delta_{k,a}^{(j)} + \text{tr} \left(\mathbf{D}_{j,a}\mathbf{G}_{k,a}\mathbf{D}_{j,a}^H \mathbf{G}_{k,a}\right) \right)} \\ & + \ds \sum_{a\in{\cal A}_j} { \ds \sum_{\substack{b\in{\cal A}_j \\ b \neq a}} {\sqrt{\eta_{j,a}^{\rm DL}} \sqrt{\eta_{j,b}^{\rm DL}}  \text{tr} \left(\mathbf{D}_{j,a}\mathbf{G}_{k,a} \right) \text{tr} \left(\mathbf{D}_{j,b}^H \mathbf{G}_{k,b} \right) }} \; .
\end{array}
\label{Interference1_1}
\end{equation}

Substituting Eq. \eqref{Interference1_1} into Eq. \eqref{Interference_all2}, and using the definitions in Section \ref{MMSE_Ch_est} we obtain

\begin{equation}
\begin{array}{llll}
&\mathbb{E}\left[ |{I}_{k,j}|^2\right] =  \ds \sum_{a\in{\cal A}_j} {\eta_{j,a}^{\rm DL} \sqrt{\eta_j} \text{tr} \left(\mathbf{G}_{j,a}\mathbf{D}_{j,a}^H \mathbf{G}_{k,a} \right) } \\ & +  \eta_k \left|\boldsymbol{\phi}_k^H \boldsymbol{\phi}_j \right|^2 \ds \sum_{a\in{\cal A}_j} \left[\eta_{j,a}^{\rm DL} \delta_{k,a}^{(j)}  \right. \\ & \left. + \ds \sum_{\substack{b\in{\cal A}_j \\ b \neq a}} {\sqrt{\eta_{j,a}^{\rm DL}} \sqrt{\eta_{j,b}^{\rm DL}} \text{tr} \left(\mathbf{D}_{j,a}\mathbf{G}_{k,a} \right) \text{tr} \left(\mathbf{D}_{j,b}^H \mathbf{G}_{k,b} \right)}\right] \; .
\end{array}
\label{Interference_all3}
\end{equation}

Finally, the downlink spectral efficiency LB in Eq. \eqref{eq:SE_MMSE_LB} can be derived by plugging Eqs. \eqref{Dk}, \eqref{Bk_final} and \eqref{Interference_all3} into Eq. \eqref{eq:SE_general}.

\end{enumerate}

\subsection{Derivation of Lemma 2}
\label{Appendix_UL}
The derivation of \eqref{eq:SE_MMSE_LB_UL} is similar to the one of  \eqref{eq:SE_MMSE_LB}. 

\begin{enumerate}

\item \textit{Computation of} $\widetilde{D}_{k}$: Using a similar procedure as in the downlink we have:

\begin{equation}
\widetilde{D}_{k}=  \sqrt{\eta_k^{\rm UL}} \ds \sum_{a\in{\cal A}_k} \ds \gamma_{k,a} \; .
\label{E_Dk_UL}
\end{equation}

\item \textit{Computation of} $\mathbb{E}\left[ \left|\widetilde{B}_{k}\right|^2\right]$: Following a similar approach as in Eqs. \eqref{Bk_final}, we obtain

\begin{equation}
\begin{array}{lll}
\mathbb{E}\left[ |\widetilde{B}_{k}|^2\right]=&\ds \eta_{k}^{\rm UL}  \sum_{a\in{\cal A}_k} \ds \left( \eta_k \widetilde{\delta}_{k,a}^{(k)} \right.\\ & \left. + \sqrt{\eta_k} \text{tr} \left(\mathbf{G}_{k,a}\mathbf{D}_{k,a}^H \mathbf{G}_{k,a}\right)-\gamma_{k,a}^2\right) \, ,
\end{array}
\label{Bk_final_UL}
\end{equation}
where $ \widetilde{\delta}_{k,a}^{(k)}$ is defined as in Eq. \eqref{delta_tilde_ka_j}.

\item \textit{Computation of} $\mathbb{E}\left[ \left|\widetilde{N}_{k}\right|^2\right]$: Since the RVs representing the noise and the wireless channel are independent, and considering that the variance of a sum of independent RVs is equal to the sum of the variances, we have
\begin{equation}
\begin{array}{llll}
&\mathbb{E}\left[ \left|\widetilde{N}_{k}\right|^2\right] = \sigma^2_w \ds \sum_{a\in{\cal A}_k} \gamma_{k,a}\; .
\end{array}
\label{E_Nk_UL}
\end{equation}

\item \textit{Computation of} $\mathbb{E}\left[ \left|\widetilde{I}_{k,j}\right|^2\right]$: Using a similar approach as in Eq. \eqref{Interference_all3}, we obtain:

\begin{equation}
\begin{array}{llll}
&\mathbb{E}\left[ |\widetilde{I}_{k,j}|^2\right] = \eta_{j}^{\rm UL} \ds \sum_{a\in{\cal A}_k} { \sqrt{\eta_k} \text{tr} \left( \mathbf{D}_{k,a}^H \mathbf{G}_{j,a} \mathbf{G}_{k,a} \right)} \\ & +  \eta_j \eta_{j}^{\rm UL} \ds \sum_{a\in{\cal A}_k} \left[ \widetilde{\delta}_{j,a}^{(k)} \right. \\ & \left. + \ds \sum_{\substack{b\in{\cal A}_k \\ b \neq a}} \text{tr} \left(\mathbf{D}_{k,a}^H \mathbf{G}_{j,a} \right)\text{tr} \left(\mathbf{D}_{k,b} \mathbf{G}_{j,b}\right)\right] \left|\boldsymbol{\phi}_j^H \boldsymbol{\phi}_k \right|^2\; .
\end{array}
\label{Interference_all3_UL}
\end{equation}

Finally,  the uplink spectral efficiency LB in Eq. \eqref{eq:SE_MMSE_LB_UL} can be derived by plugging Eqs. \eqref{E_Dk_UL}, \eqref{Bk_final_UL}, \eqref{E_Nk_UL}, and  \eqref{Interference_all3_UL} in Eq. \eqref{eq:SE_general_UL}.

\end{enumerate}

\subsection{Derivation of Lemma 3}
\label{Appendix_Lemma3}

To circumvent the challenge of non-differentiability of the Problem \eqref{Prob:MinRate}, we reformulate it as
\begin{subequations}\label{Prob:MinRate3}
\begin{align}
&\ds\max_{\overline{\bbeta},t}\;t \label{Prob:aMinRate3}\\
&\;\textrm{s.t.}\; \sum_{k\in{\cal K}_a}\overline{\eta}_{k,a}^{\rm DL} \rho_{a,k}^{\rm DL} \gamma_{k,a} \leq \eta^{\rm DL}_a\;,\forall\;a=1,\ldots,N_A\label{Prob:bMinRate3}\\
&\;\;\;\quad \overline{\eta}_{k,a}^{\rm DL}\geq 0\;,\forall\;a=1,\ldots,N_A,\;k=1,\ldots,K,\label{Prob:cMinRate3} \\
&\;\;\;\quad \mathcal{R}_{k}^{\rm DL}(\overline{\bbeta}) \geq t , \; \forall \; k=1,\ldots, K.
\label{Prob:dMinRate3}
\end{align}
\end{subequations}

To solve \eqref{Prob:MinRate3}, following an approach similar to that in  \cite{BuzziZappone_PIMRC2017}, 
the framework of successive lower-bound maximization \cite{RazaviyaynSIAM},  which combines the tools of alternating optimization \cite[Section 2.7]{BertsekasNonLinear} and sequential convex programming \cite{SeqCvxProg78} can be used. 
In particular, consider Problem \eqref{Prob:MinRate3} and define the $L$-dimensional variable blocks $\overline{\bbeta}_{a}^{(p)}$, $a=1,\ldots,N_A, \; p=1,\ldots,P_a$,  collecting the $p$-th block of ``normalized'' transmit powers of the $a$-th AP. Then,  the minimum-rate maximization problem can be reformulated as  
\begin{subequations}\label{Prob:MinRateSub}
\begin{align}
&\ds\max_{\overline{\bbeta}_{a}^{(p)},t}\; t \label{Prob:aMinRateSub}\\
&\;\textrm{s.t.}\; \sum_{k\in{\cal K}_a}\overline{\eta}_{k,a}^{\rm DL} \rho_{a,k}^{\rm DL} \gamma_{k,a} \leq\eta^{\rm DL}_a\;,\forall\;a=1,\ldots,N_A\label{Prob:bMinRateSub}\\
&\;\;\;\quad \overline{\eta}_{k,a}^{\rm DL}\geq 0\;,\forall\;a=1,\ldots,N_A,\;k=1,\ldots,K.\label{Prob:cMinRateSub}\\
&\;\;\;\quad \mathcal{R}_{k}^{\rm DL}(\overline{\bbeta}_{a}^{(p)},\overline{\bbeta}_{-a}^{(-p)}) \geq t , \; \forall \; k=1,\ldots, K.
\label{Prob:dMinRateSub}
\end{align}
\end{subequations}
Although \eqref{Prob:MinRateSub} is still non-convex, its complexity is significantly lower than that of \eqref{Prob:MinRate3}, since only the block of $L$ transmit powers of the $a$-th AP are being optimized. Notice that the $k$-th user's downlink achievable rate can be written  as 
\begin{equation}
\mathcal{R}_k^{\rm DL}(\overline{\bbeta})=W \ds\frac{\tau_{\rm d}}{\tau_c} \log_2  \left[ g_1\left(\overline{\bbeta}\right)+g_2\left(\overline{\bbeta}\right)\right] - W \ds\frac{\tau_{\rm d}}{\tau_c} \log_2  \left[g_2\left(\overline{\bbeta}\right)\right]\, ,
\label{Eq:Ratek_diff}
\end{equation}
where $g_1\left(\overline{\bbeta}\right)$ and $g_2\left(\overline{\bbeta}\right)$ are defined as
\begin{equation}
\begin{array}{llll}
&g_1\left(\overline{\bbeta}\right)=\left(\ds \sum_{a\in{\cal A}_k} {\ds \sqrt{\overline{\eta}_{k,a}^{\rm DL}\rho_{a,k}^{\rm DL}} \gamma_{k,a}} \right)^2 
\;,
\\ &g_2\left(\overline{\bbeta}\right)= \ds \sum_{j \in \mathcal{K}} \sqrt{\eta_j} \ds \sum_{a\in{\cal A}_j} \overline{\eta}_{j,a}^{\rm DL} \rho_{a,k}^{\rm DL} \text{tr} \left( \mathbf{G}_{j,a} \mathbf{D}_{j,a}^H \mathbf{G}_{k,a} \right)  + \sigma^2_z \\&  + \ds \sum_{j \in \mathcal{K}\backslash k} \eta_k \left|\boldsymbol{\phi}_k^H \boldsymbol{\phi}_j \right|^2 \left\lbrace \ds \sum_{a\in{\cal A}_j} \left[\overline{\eta}_{j,a}^{\rm DL} \rho_{a,k}^{\rm DL}\delta_{k,a}^{(j)} \right. \right. \\ & \left. \left. + \ds \sum_{\substack{b\in{\cal A}_j \\ b \neq a}} \sqrt{\overline{\eta}_{j,a}^{\rm DL} \rho_{a,k}^{\rm DL}} \sqrt{\overline{\eta}_{j,b}^{\rm DL}\rho_b^{\rm DL}} \text{tr} \left(\mathbf{D}_{j,a}\mathbf{G}_{k,a}\right)\text{tr} \left(\mathbf{D}_{j,b}^H\mathbf{G}_{k,b}\right)\right]\right\rbrace.
\end{array}\;
\label{g1_g2}
\end{equation}
Since the function in Eq. \eqref{Eq:Ratek_diff} is non-concave, even with respect to only the variable block $\overline{\bbeta}_{a}^{(p)}$, optimization \eqref{Prob:MinRateSub} will be tackled by sequential optimization. To this end, we need a lower-bound of $\mathcal{R}_k^{\rm DL}(\overline{\bbeta})$, which fulfills properties in \cite{SeqCvxProg78}, while at the same time leading to a simple optimization problem. Using the fact that the function $f(x,y)=\sqrt{xy}$ is jointly concave in $x$ and $y$, for $x,y>0$, and since the function $\log_2(\cdot)$ is an increasing function, and  summation preserves concavity, the rate function in \eqref{Eq:Ratek_diff} is the difference of two concave functions \cite{BuzziZappone_PIMRC2017,AloBuZap5GWF2018}. Recalling that any concave function is upper-bounded by its Taylor expansion around any given point $\overline{\bbeta}_{a,0}^{(p)}$, a concave lower-bound of $\mathcal{R}_k^{\rm DL}(\overline{\bbeta})$ is obtained as
\begin{equation}
\begin{array}{llll}
&\mathcal{R}_k^{\rm DL}\left(\overline{\bbeta}_{a}^{(p)}\right)  \geq \widetilde{\mathcal{R}}_k^{\rm DL}\left(\overline{\bbeta}_{a}^{(p)}\right)= \\ & W \ds\frac{\tau_{\rm d}}{\tau_c} \log_2  \left[ g_1\left(\overline{\bbeta}_{a}^{(p)}\right)+g_2\left(\overline{\bbeta}_{a}^{(p)}\right)\right] - W \ds\frac{\tau_{\rm d}}{\tau_c} \log_2  \left[g_2\left(\overline{\bbeta}_{a,0}^{(p)}\right)\right] \\ & -
 W \ds\frac{\tau_{\rm d}}{\tau_c} \nabla_{\overline{\bbeta}_{a}^{(p)}}^T  \log_2 \left[ g_2\rvert_{\overline{\bbeta}_{a,0}^{(p)}}\left(\overline{\bbeta}_{a}^{(p)}\right)\right]\, \left( \overline{\bbeta}_{a}^{(p)} - \overline{\bbeta}_{a,0}^{(p)} \right).
\end{array}
\label{Eq:Ratek_diff_lower_bound}
\end{equation}
Relying on this bound, we can solve Problem \eqref{Prob:MinRateSub} through the sequential optimization method, by defining the $p$-th problem of the sequence, ${\cal P}_{p}$, as the convex optimization problem in \eqref{Prob:MinRateApp}. Following the successive lower-bound maximization framework, it thus follows that the original optimization problem \eqref{Prob:MinRate} can be solved using the procedure in Algorithm 1.

%

\ifCLASSOPTIONcaptionsoff
  \newpage
   \fi
\bibliographystyle{IEEEtran}
\bibliography{Strings_Gio,Bib_Gio,References}

\begin{IEEEbiographynophoto}{Carmen D'Andrea} (S'18 - M'20) was born in Caserta, Italy on 16 July 1991. She received the B.S. and M.S. degrees, both with honors, in Telecommunications Engineering from University of Cassino and Lazio Meridionale in 2013 and 2015, respectively. In 2017, she was a Visiting Ph.D. student with the Wireless Communications (WiCom) Research Group 
in the Department of Information and Communication Technologies at Universitat Pompeu Fabra in Barcelona, Spain. 
In 2019, she received the Ph.D. degree with highest marks in Electrical and Information Engineering from 
University of Cassino and Lazio Meridionale. She is currently a Post-Doctoral Researcher with Department of Electrical and Information Engineering, University of Cassino and Lazio Meridionale. Her research interests are focused on wireless communication and signal processing, with a current emphasis on mmWave communications and massive MIMO systems, in both colocated and distributed setups.
\end{IEEEbiographynophoto}
\vspace{-2cm}
\begin{IEEEbiographynophoto}{Adrian Garcia-Rodriguez} is a Research Scientist in Nokia Bell Labs (Ireland), where he focuses on the design of UAV communications and next-generation 802.11 technologies. He joined Bell Labs in 2016, after receiving the Ph.D. degree in Electrical and Electronic Engineering from University College London (U.K.). Adrian is a co-inventor of fifteen filed patent families and co-author of 40+ IEEE publications. He was the recipient of the Best Paper Award in PIMRC’19 and was named an Exemplary Reviewer for IEEE Commun. Letters in 2016, and both IEEE Trans. on Wireless Commun. and IEEE Trans. on Commun. in 2017.
\end{IEEEbiographynophoto}
\vspace{-2cm}
\begin{IEEEbiographynophoto}{Giovanni Geraci} is an Assistant Professor and Junior Leader Fellow at UPF Barcelona (Spain). He earned a Ph.D. from the UNSW Sydney (Australia) in 2014, and was a Research Scientist with Nokia Bell Labs (Ireland) in 2016-2018. His background also features research appointments at SUTD (Singapore) in 2014-2015, UT Austin (USA) in 2013, CentraleSupélec (France) in 2012, and Alcatel-Lucent (Italy) in 2009. He has been serving as an Editor for the IEEE Transactions on Wireless Communications and IEEE Communications Letters, and as a workshop co-chair at IEEE ICC, IEEE Globecom, and Asilomar. He has been a panelist, workshop keynote, and industrial seminar or tutorial speaker at IEEE ICC, IEEE Globecom, IEEE WCNC, IEEE PIMRC, and IEEE VTC Spring. He has co-authored 50+ IEEE publications with 1500+ citations, and is co-inventor of a dozen filed patent families. Giovanni was the recipient of the Best Paper Award at IEEE PIMRC'19 and of the IEEE ComSoc Outstanding Young Researcher Award for Europe, Middle-East \& Africa 2018.
\end{IEEEbiographynophoto}
\vspace{-2cm}
\begin{IEEEbiographynophoto}{Lorenzo Galati Giordano} (M'15) is Member of Technical Staff at Nokia Bell Labs Ireland since 2015. Lorenzo received the M.Sc. and the Ph.D. degrees in wireless communication from Politecnico di Milano, Italy, in 2005 and 2010, respectively, and the master's degree in Innovation Management from IlSole24Ore Business School, Italy, in 2014. He was also Marie-Curie Short Term Fellow at University of Bedfordshire (UK) in 2008, researcher associate with the Italian National Research Council in 2010 and R\&D Engineer for Azcom Technology, an Italian SME, from 2010 to 2014.  Lorenzo has more than 10 years of academical and industrial research experience on wireless communication systems and protocols, holds commercial patents and publications in prestigious IEEE journals and conferences. During the past years, Lorenzo contributed to the Nokia F-Cell project, an innovative self-powered and auto-connected drone deployed small cell served by massive MIMO wireless backhaul, which received the CTIA Emerging Technology 2016 Award. Lorenzo’s current focus is on future indoor networks and next generation Wi-Fi technologies, an area where he is contributing with large antenna arrays solutions for the unlicensed spectrum. 
\end{IEEEbiographynophoto}
\vspace{-2cm}
\begin{IEEEbiographynophoto}{Stefano Buzzi} (M'98-SM'07) is Full Professor at the University of Cassino and Lazio Meridionale, Italy. 
He received the M.Sc. degree (summa cum laude) in Electronic Engineering in 1994, and the Ph.D. degree in Electrical and 
Computer Engineering in 1999, both from the University of Naples “Federico II”. He has had short-term research appointments 
at Princeton University, Princeton (NJ), USA in 1999, 2000, 2001 and 2006. He is a former Associate Editor of the \textit{IEEE Signal 
Processing Letters} and of the \textit{IEEE Communications Letters}, has been the lead guest editor of three \textit{IEEE JSAC} special issues 
(June 2014, April 2016, and April 2019), while is currently serving as an Editor for the \textit{IEEE Transactions on Wireless Communications}. 
He is also a Member of the IEEE Future Networks Editorial Board, and serves regularly as TPC member of several international
conferences. Dr. Buzzi’s research interests are in the broad field of 
communications and signal processing, with emphasis on wireless communications. He has co-authored about 
160 technical peer-reviewed journal and conference papers, and among these, the highly-cited survey paper “What will 5G be?” 
(IEEE JSAC, June 2014) on 5G wireless networks.
\end{IEEEbiographynophoto}

\end{document}